# Set Theory for Verification: II

*Induction and Recursion*

Lawrence C. Paulson
*Computer Laboratory, University of Cambridge*

**Abstract.** A theory of recursive definitions has been mechanized in Isabelle's Zermelo-Fraenkel (ZF) set theory. The objective is to support the formalization of particular recursive definitions for use in verification, semantics proofs and other computational reasoning.

*Inductively defined sets* are expressed as least fixedpoints, applying the Knaster-Tarski Theorem over a suitable set. *Recursive functions* are defined by well-founded recursion and its derivatives, such as transfinite recursion. *Recursive data structures* are expressed by applying the Knaster-Tarski Theorem to a set, such as $V_\omega$, that is closed under Cartesian product and disjoint sum.

Worked examples include the transitive closure of a relation, lists, variable-branching trees and mutually recursive trees and forests. The Schröder-Bernstein Theorem and the soundness of propositional logic are proved in Isabelle sessions.

**Key words:** Isabelle, set theory, recursive definitions, the Schröder-Bernstein Theorem

## Contents

## 1. Introduction

Recursive definitions pervade theoretical Computer Science. Part I of this work [22] has described the mechanization of a theory of functions within Zermelo-Fraenkel (ZF) set theory using the theorem prover Isabelle. Part II develops a mechanized theory of recursion for ZF: least fixedpoints, recursive functions and recursive data structures. Particular instances of these can be generated rapidly, to support verifications and other computational proofs in ZF set theory.

The importance of this theory lies in its relevance to automation. I describe the Isabelle proofs in detail, so that they can be reproduced in other set theory provers. It also serves as an extended demonstration of how mathematics is developed using Isabelle. Two Isabelle proofs are presented: the Schröder-Bernstein Theorem and a soundness theorem for propositional logic.

### 1.1. Outline of the Paper

Part I [22] contains introductions to axiomatic set theory and Isabelle. Part II, which is the present document, proceeds as follows.

- Section 2 presents a treatment of least fixedpoints based upon the Knaster-Tarski Theorem. Examples include transitive closure and the Schröder-Bernstein Theorem.
- Section 3 treats recursive functions. It includes a detailed derivation of well-founded recursion. The ordinals, $\in$-recursion and the cumulative hierarchy are defined in order to derive a general recursion operator for recursive data structures.
- Section 4 treats recursive data structures, including mutual recursion. It presents examples of various types of lists and trees. Little new theory is required.
- Section 5 is a case study to demonstrate all of the techniques. It describes an Isabelle proof of the soundness and completeness of propositional logic.
- Section 6 outlines related work and draws brief conclusions.

### 1.2. Preliminary Definitions

For later reference, I summarize below some concepts defined in Part I [22], mainly in §7.5. Ideally, you should read the whole of Part I before continuing.

A *binary relation* is a set of ordered pairs. Isabelle's set theory defines the usual operations: converse, domain, range, etc. The infix operator `` denotes image.

$$\langle y,x\rangle \in \texttt{converse}(r) \;\leftrightarrow\; \langle x,y\rangle \in r$$

$$
\begin{aligned}
x \in \texttt{domain}(r) &\leftrightarrow \exists y \,.\, \langle x, y\rangle \in r \\
y \in \texttt{range}(r) &\leftrightarrow \exists x \,.\, \langle x, y\rangle \in r \\
\texttt{field}(r) &\equiv \texttt{domain}(r) \cup \texttt{range}(r) \\
y \in (r \text{ “ } A) &\leftrightarrow \exists_{x\in A} \,.\, \langle x, y\rangle \in r
\end{aligned}
$$

The *definite description* operator $\iota x \,.\, \psi(x)$ denotes the unique $a$ satisfying $\psi(a)$, if such exists. See §7.2 of Part I for its definition and discussion.

*Functions* are single-valued binary relations. Application and $\lambda$-abstraction are defined as follows:

$$
\begin{aligned}
f\text{‘}a &\equiv \iota y \,.\, \langle a, y\rangle \in f \\
\lambda_{x\in A} \,.\, b(x) &\equiv \{\langle x, b(x)\rangle \,.\, x \in A\}
\end{aligned}
$$

## 2. Least Fixedpoints

One aspect of the Isabelle ZF theory of recursion concerns sets defined by least fixedpoints. I use an old result, the Knaster-Tarski Theorem. A typical application is to formalize the set of theorems inductively defined by a system of inference rules. The set being defined must be a subset of another set already available. Later (§4.2) we shall construct sets large enough to contain various recursive data structures, which can be ‘carved out’ using the Knaster-Tarski Theorem.

This section gives the Isabelle formulation of the Theorem. The least fixedpoint satisfies a general induction principle that can be specialized to obtain structural induction rules for the natural numbers, lists and trees. The transitive closure of a relation is defined as a least fixedpoint and its properties are proved by induction. A least fixedpoint argument also yields a simple proof of the Schröder-Bernstein Theorem. Part of this proof is given in an interactive session, to demonstrate Isabelle's ability to synthesize terms.

### 2.1. The Knaster-Tarski Theorem

The Knaster-Tarski Theorem states that every monotone function over a complete lattice has a fixedpoint. (Davey and Priestley discuss and prove the Theorem [7].) Usually a greatest fixedpoint is exhibited, but a dual argument yields the least fixedpoint.

A partially ordered set $P$ is a *complete lattice* if, for every subset $S$ of $P$, the least upper bound and greatest lower bound of $S$ are elements of $P$. In Isabelle's implementation of ZF set theory, the theorem is proved for a special case: powerset lattices of the form $\wp(D)$, for a set $D$. The partial ordering is $\subseteq$; upper bounds are unions; lower bounds are intersections.

Other complete lattices could be useful. Mutual recursion can be expressed as a fixedpoint in the lattice $\wp(D_1) \times \cdots \times \wp(D_n)$, whose elements are $n$-tuples, with a component-wise ordering. But proving the Knaster-Tarski Theorem in its full generality would require a cumbersome formalization of complete lattices. The Isabelle ZF treatment of mutual recursion uses instead the lattice $\wp(D_1+\cdots+D_n)$, which is order-isomorphic[1] to $\wp(D_1)\times\cdots\times\wp(D_n)$.

The predicate $\texttt{bnd_mono}(D,h)$ expresses that $h$ is monotonic and bounded by $D$, while $\texttt{lfp}(D,h)$ denotes $h$'s least fixedpoint, a subset of $D$:

$$\begin{aligned}\texttt{bnd_mono}(D,h) &\equiv h(D)\subseteq D \land (\forall x\, y\, .\, x\subseteq y \land y\subseteq D \to h(x)\subseteq h(y))\\ \texttt{lfp}(D,h) &\equiv \bigcap\{X\in\wp(D)\,.\,h(X)\subseteq X\}\end{aligned}$$

These are binding operators; in Isabelle terminology, $h$ is a meta-level function. I originally defined lfp for object-level functions, but this needlessly complicated proofs. A function in set theory is a set of pairs. There is an obvious correspondence between meta- and object-level functions with domain $\wp(D)$, associating $h$ with $\lambda_{X\in\wp(D)}\,.\,h(X)$. The latter is an element of the set $\wp(D)\to\wp(D)$, but this is irrelevant to the theorem at hand. What matters is the mapping from $X$ to $h(X)$.

Virtually all the functions in this paper are meta-level functions, not sets of pairs. One exception is in the well-founded recursion theorem below (§3.1), where the construction of the recursive function simply must be regarded as the construction of a set of pairs. Object-level functions stand out because they require an application operator: we must write $f`x$ instead of $f(x)$.

The Isabelle theory derives rules asserting that $\texttt{lfp}(D,h)$ is the least prefixedpoint of $h$, and (if $h$ is monotonic) a fixedpoint:

$$\frac{h(A)\subseteq A \quad A\subseteq D}{\texttt{lfp}(D,h)\subseteq A} \qquad \frac{\texttt{bnd_mono}(D,h)}{\texttt{lfp}(D,h)=h(\texttt{lfp}(D,h))}$$

The second rule above is one form of the Knaster-Tarski Theorem. Another form of the Theorem constructs a greatest fixedpoint; this justifies coinductive definitions [23], but will not concern us here.

## 2.2. The Bounding Set

When justifying some instance of $\texttt{lfp}(D,h)$, showing that $h$ is monotone is generally easy, if it is true at all. Harder is to exhibit a *bounding set*, namely some $D$ satisfying $h(D)\subseteq D$. Much of the work reported below involves constructing bounding sets for use in fixedpoint definitions. Let us consider some examples.

- *The natural numbers.* The Axiom of Infinity (see §3.3) asserts that there is a bounding set Inf for the mapping $\lambda X\,.\,\{0\}\cup\{\texttt{succ}(i)\,.\,i\in X\}$. This justifies defining the set of natural numbers by

  $$\texttt{nat} \equiv \texttt{lfp}(\texttt{Inf},\ \lambda X\,.\,\{0\}\cup\{\texttt{succ}(i)\,.\,i\in X\}).$$

– *Lists and trees.* Let $A+B$ denote the disjoint sum of the sets $A$ and $B$ (defined below in §4.1). Consider defining the set of lists over $A$, satisfying the recursion equation
$$\texttt{list}(A) = \{\emptyset\} + A \times \texttt{list}(A).$$
This requires a set closed under the mapping $\lambda X . \{\emptyset\} + A \times X$. Section 4.2 defines a set $\texttt{univ}(A)$ with useful closure properties:
$$\begin{array}{rl} A \subseteq \texttt{univ}(A) & \texttt{univ}(A) \times \texttt{univ}(A) \subseteq \texttt{univ}(A) \\ \texttt{nat} \subseteq \texttt{univ}(A) & \texttt{univ}(A) + \texttt{univ}(A) \subseteq \texttt{univ}(A) \end{array}$$
This set contains all finitely branching trees over $A$, and will allow us to define a wide variety of recursive data structures.

– The Isabelle ZF theory also constructs bounding sets for *infinitely branching trees.*

– The *powerset operator* is monotone, but has no bounding set. Cantor's Theorem implies that there is no set $D$ such that $\wp(D) \subseteq D$.

### 2.3. A General Induction Rule

Because $\texttt{lfp}(D, h)$ is a least fixedpoint, it enjoys an induction rule. Consider the set of natural numbers, $\texttt{nat}$. Suppose $\psi(0)$ holds and that $\psi(x)$ implies $\psi(\texttt{succ}(x))$ for all $x \in \texttt{nat}$. Then the set $\{x \in \texttt{nat} . \psi(x)\}$ contains 0 and is closed under successors. Because $\texttt{nat}$ is the least such set, we obtain $\texttt{nat} \subseteq \{x \in \texttt{nat} . \psi(x)\}$. Thus, $x \in \texttt{nat}$ implies $\psi(x)$.

To derive an induction rule for an arbitrary least fixedpoint, the chief problem is to express the rule's premises. Suppose we have defined $A \equiv \texttt{lfp}(D, h)$ and have proved $\texttt{bnd_mono}(D, h)$. Define the set
$$A_\psi \equiv \{x \in A . \psi(x)\}.$$
Now suppose $x \in h(A_\psi)$ implies $\psi(x)$ for all $x$. Then $h(A_\psi) \subseteq A_\psi$ and we conclude $A \subseteq A_\psi$. This derives the general induction rule
$$\frac{A \equiv \texttt{lfp}(D, h) \quad a \in A \quad \texttt{bnd_mono}(D, h) \quad \begin{array}{c}[x \in h(A_\psi)]_x \\ \vdots \\ \psi(x)\end{array}}{\psi(a)}$$
The last premise states the closure properties of $\psi$, normally expressed as separate 'base cases' and 'induction steps.' (As in Part I of this paper, the subscripted variable in the assumption stands for a proviso on the rule: $x$ must not be free in the conclusion or other assumptions.)

To demonstrate this rule, consider again the natural numbers. The appropriate $h$ satisfies
$$h(\texttt{nat}_\psi) = \{0\} \cup \{\texttt{succ}(i) . i \in \texttt{nat}_\psi\}.$$

Now $x \in h(\mathtt{nat}_\psi)$ if and only if $x = 0$ or $x = \mathtt{succ}(i)$ for some $i \in \mathtt{nat}$ such that $\psi(i)$. We may instantiate the rule above to

$$\frac{n \in \mathtt{nat} \qquad \begin{matrix}[x \in h(\mathtt{nat}_\psi)]_x \\ \vdots \\ \psi(x)\end{matrix}}{\psi(n)}$$

and quickly derive the usual induction rule

$$\frac{n \in \mathtt{nat} \quad \psi(0) \qquad \begin{matrix}[x \in \mathtt{nat} \quad \psi(x)]_x \\ \vdots \\ \psi(\mathtt{succ}(x))\end{matrix}}{\psi(n)}$$

## 2.4. Monotonicity

The set $\mathtt{lfp}(D, h)$ is a fixedpoint if $h$ is monotonic. The Isabelle ZF theory derives many rules for proving monotonicity; Isabelle's classical reasoner proves most of them automatically. Here are the rules for union and product:

$$\frac{A \subseteq C \quad B \subseteq D}{A \cup B \subseteq C \cup D} \qquad \frac{A \subseteq C \quad B \subseteq D}{A \times B \subseteq C \times D}$$

Here are the rules for set difference and image:

$$\frac{A \subseteq C \quad D \subseteq B}{A - B \subseteq C - D} \qquad \frac{r \subseteq s \quad A \subseteq B}{r``A \subseteq s``B}$$

And here is the rule for general union:

$$\frac{A \subseteq C \quad \begin{matrix}[x \in A]_x \\ \vdots \\ B(x) \subseteq D(x)\end{matrix}}{(\bigcup_{x \in A} . B(x)) \subseteq (\bigcup_{x \in C} . D(x))}$$

There is even a rule that $\mathtt{lfp}$ is itself monotonic.[2] This justifies nested applications of $\mathtt{lfp}$:

$$\frac{\mathtt{bnd_mono}(D, h) \quad \mathtt{bnd_mono}(E, i) \quad \begin{matrix}[X \subseteq D]_X \\ \vdots \\ h(X) \subseteq i(X)\end{matrix}}{\mathtt{lfp}(D, h) \subseteq \mathtt{lfp}(E, i)}$$

### 2.5. Application: Transitive Closure of a Relation

Let $\mathtt{id}(A)$ denote the identity relation on $A$, namely $\{\langle x,x\rangle \,.\, x \in A\}$. Then the reflexive/transitive closure $r^*$ of a relation $r$ may be defined as a least fixedpoint:

$$r^* \equiv \mathtt{lfp}(\mathtt{field}(r)\times\mathtt{field}(r),\ \lambda s\,.\,\mathtt{id}(\mathtt{field}(r))\cup(r\circ s))$$

The mapping $\lambda s\,.\,\mathtt{id}(\mathtt{field}(r))\cup(r\circ s)$ is monotonic and bounded by $\mathtt{field}(r)\times\mathtt{field}(r)$, by virtue of similar properties for union and composition. The Knaster-Tarski Theorem yields

$$r^* = \mathtt{id}(\mathtt{field}(r))\cup(r\circ r^*).$$

This recursion equation affords easy proofs of the introduction rules for $r^*$:

$$\frac{a\in\mathtt{field}(r)}{\langle a,a\rangle\in r^*}\qquad\frac{\langle a,b\rangle\in r^*\quad\langle b,c\rangle\in r}{\langle a,c\rangle\in r^*}$$

Because $r^*$ is recursively defined, it admits reasoning by induction. Using the general induction rule for $\mathtt{lfp}$, the following rule can be derived simply:

$$\frac{\langle a,b\rangle\in r^*\qquad\begin{matrix}[x\in\mathtt{field}(r)]_x\\ \vdots\\ \psi(\langle x,x\rangle)\end{matrix}\qquad\begin{matrix}[\psi(\langle x,y\rangle)\quad\langle x,y\rangle\in r^*\quad\langle y,z\rangle\in r]_{x,y,z}\\ \vdots\\ \psi(\langle x,z\rangle)\end{matrix}}{\psi(\langle a,b\rangle)}\qquad(1)$$

This is the natural elimination rule for $r^*$ because its minor premises reflect the form of its introduction rules [25]; it is however cumbersome. A simpler rule starts from the idea that if $\langle a,b\rangle\in r^*$ then there exist $a_0$, $a_1$, …, $a_n$ such that (writing $r$ as an infix relation)

$$a = a_0\,r\,a_1\,r\,\cdots\,r\,a_n = b.$$

If $\psi$ holds at $a$ and is preserved by $r$, then $\psi$ must hold at $b$:

$$\frac{\langle a,b\rangle\in r^*\quad\psi(a)\qquad\begin{matrix}[\psi(y)\quad\langle a,y\rangle\in r^*\quad\langle y,z\rangle\in r]_{y,z}\\ \vdots\\ \psi(z)\end{matrix}}{\psi(b)}\qquad(2)$$

Formally, the rule follows by assuming its premises and instantiating the original induction rule (1) with the formula $\psi'(z)$, where

$$\psi'(z) \equiv \forall w\,.\,z=\langle a,w\rangle\to\psi(w).$$

Reasoning about injectivity of ordered pairing, we eventually derive

$$\forall w\,.\,\langle a,b\rangle=\langle a,w\rangle\to\psi(w)$$

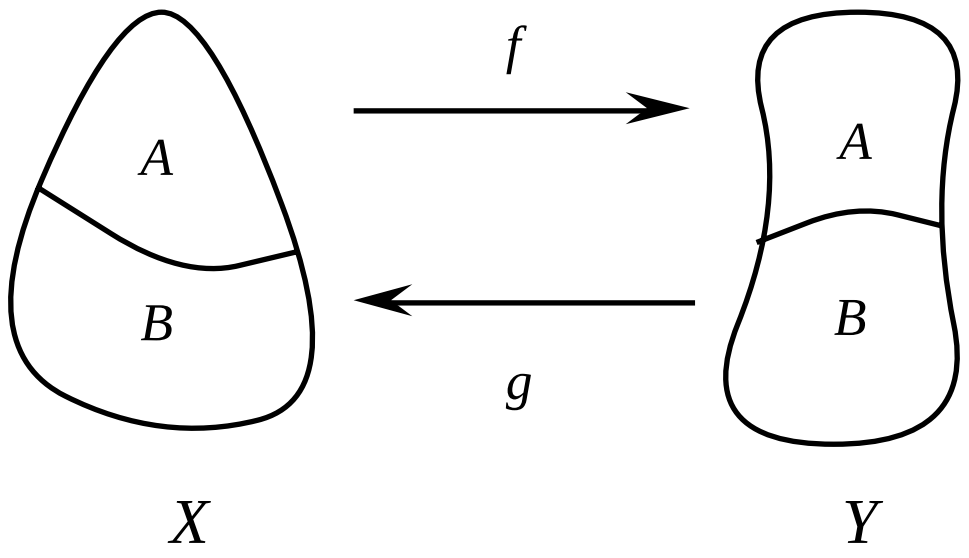

*Figure 1.* Banach's Decomposition Theorem

and reach the conclusion, $\psi(b)$.

To demonstrate the simpler induction rule (2), let us show that $r^*$ is transitive. Here is a concise proof of $\langle c, b\rangle \in r^*$ from the assumptions $\langle c, a\rangle \in r^*$ and $\langle a, b\rangle \in r^*$:

$$\frac{\langle a,b\rangle \in r^* \quad \langle c,a\rangle \in r^* \quad \dfrac{\langle c,y\rangle \in r^* \quad \langle y,z\rangle \in r}{\langle c,z\rangle \in r^*}}{\langle c,b\rangle \in r^*}$$

The transitive closure $r^+$ of a relation $r$ is defined by $r^+ \equiv r \circ r^*$ and its usual properties follow immediately.

## 2.6. Application: The Schröder-Bernstein Theorem

The Schröder-Bernstein Theorem plays a vital role in the theory of cardinal numbers. If there are two injections $f : X \to Y$ and $g : Y \to X$, then the Theorem states that there is a bijection $h : X \to Y$. Halmos [11] gives a direct but complicated proof. Simpler is to use the Knaster-Tarski Theorem to prove a key lemma, Banach's Decomposition Theorem [7].

Recall from §1.2 the image and converse operators. These apply to functions also, because functions are relations in set theory. If $f$ is an injection then $\texttt{converse}(f)$ is a function, conventionally written $f^{-1}$. Write $f \restriction A$ for the restriction of function $f$ to the set $A$, defined by

$$f \restriction A \equiv \lambda_{x\in A} \,.\, f`x$$

### 2.6.1. *The Informal Proof*

Suppose $f : X \to Y$ and $g : Y \to X$ are functions. Banach's Decomposition Theorem states that both $X$ and $Y$ can be partitioned (see Figure 1) into regions $A$ and $B$, satisfying six equations:

$$\begin{array}{lll} X_A \cap X_B = \emptyset & X_A \cup X_B = X & f``X_A = Y_A \\ Y_A \cap Y_B = \emptyset & Y_A \cup Y_B = Y & g``Y_B = X_B \end{array}$$

To prove Banach's Theorem, define

$$
\begin{aligned}
X_A &\equiv \texttt{lfp}(X,\ \lambda W.\, X - g``(Y - f``W)) \\
X_B &\equiv X - X_A \\
Y_A &\equiv f``X_A \\
Y_B &\equiv Y - Y_A
\end{aligned}
$$

Five of the six equations follow at once. The mapping in `lfp` is monotonic and yields a subset of $X$. Thus Tarski's Theorem yields $X_A = X - g``(Y - f``X_A)$, which justifies the last equation:

$$
\begin{aligned}
g``Y_B &= g``(Y - f``X_A) \\
&= X - (X - g``(Y - f``X_A)) \\
&= X - X_A \\
&= X_B.
\end{aligned}
$$

To prove the Schröder-Bernstein Theorem, let $f$ and $g$ be injections (for the Banach Theorem, they only have to be functions). Partition $X$ and $Y$ as above. The desired bijection between $X$ and $Y$ is $(f \restriction X_A) \cup (g \restriction Y_B)^{-1}$.

## 2.7. Proving the Schröder-Bernstein Theorem in Isabelle

This section sketches the Isabelle proof of the Schröder-Bernstein Theorem; Isabelle synthesizes the bijection automatically. See Part I for an overview of Isabelle [22, §2]. As usual, the proofs are done in small steps in order to demonstrate Isabelle's workings.

### 2.7.1. *Preliminaries for Banach's Decomposition Theorem*

Most of the work involves proving Banach's Theorem. First, we establish monotonicity of the map supplied to `lfp`:

```
bnd_mono(X, %W. X - g``(Y - f``W))
```

The proof is trivial, and omitted; the theorem is stored as `decomp_bnd_mono`.

Next, we prove the last equation in Banach's Theorem:

```
val [gfun] = goal Cardinal.thy
    "g: Y->X ==>                                        \
\    g``(Y - f`` lfp(X, %W. X - g``(Y - f``W))) =       \
\    X - lfp(X, %W. X - g``(Y - f``W)) ";
```

Isabelle responds by printing an initial proof state consisting of one subgoal, the equation to be proved.

```
Level 0
g `` (Y - f `` lfp(X,%W. X - g `` (Y - f `` W))) =
X - lfp(X,%W. X - g `` (Y - f `` W))
 1. g `` (Y - f `` lfp(X,%W. X - g `` (Y - f `` W))) =
    X - lfp(X,%W. X - g `` (Y - f `` W))
```

The first step is to use monotonicity and Tarski's Theorem to substitute for $\texttt{lfp}(\cdots)$. Unfortunately, there are two occurrences of $\texttt{lfp}(\cdots)$ and the substitution must unfold only the second one. The relevant theorems are combined and then instantiated with a template specifying where the substitution may occur.

```
by (res_inst_tac [("P", "%u. ?v = X-u")]
     (decomp_bnd_mono RS lfp_Tarski RS ssubst) 1);
```

```
Level 1
g `` (Y - f `` lfp(X,%W. X - g `` (Y - f `` W))) =
X - lfp(X,%W. X - g `` (Y - f `` W))
 1. g `` (Y - f `` lfp(X,%W. X - g `` (Y - f `` W))) =
    X - (X - g `` (Y - f `` lfp(X,%W. X - g `` (Y - f `` W))))
```

Observe the substitution's effect upon subgoal 1. We now invoke Isabelle's simplifier, supplying basic facts about subsets, complements, functions and images. This simplifies $X - (X - g``(Y - f``\texttt{lfp}(\cdots)))$ to $g``(Y - f``\texttt{lfp}(\cdots))$, which proves the subgoal.

```
by (simp_tac
    (ZF_ss addsimps [subset_refl, double_complement, Diff_subset,
                     gfun RS fun_is_rel RS image_subset]) 1);
```

```
Level 2
g `` (Y - f `` lfp(X,%W. X - g `` (Y - f `` W))) =
X - lfp(X,%W. X - g `` (Y - f `` W))
No subgoals!
```

The proof is finished. We name this theorem for later reference during the proof session.

```
val Banach_last_equation = result();
```

2.7.2. *The Proof of Banach's Decomposition Theorem*

We are now ready to prove Banach's Theorem proper:

```
val prems = goal Cardinal.thy
    "[| f: X->Y;  g: Y->X |] ==>                              \
\    EX XA XB YA YB. (XA Int XB = 0) & (XA Un XB = X) &    \
\                    (YA Int YB = 0) & (YA Un YB = Y) &    \
\                    f``XA=YA & g``YB=XB";
```

```
Level 0
EX XA XB YA YB.
   XA Int XB = 0 &
   XA Un XB = X &
   YA Int YB = 0 & YA Un YB = Y & f `` XA = YA & g `` YB = XB

 1. EX XA XB YA YB.
       XA Int XB = 0 &
       XA Un XB = X &
       YA Int YB = 0 & YA Un YB = Y & f `` XA = YA & g `` YB = XB
```

Starting in the initial proof state, we apply a command to strip the existential quantifiers and conjunctions repeatedly. The result is a proof state consisting of six subgoals:

```
by (REPEAT (FIRSTGOAL (resolve_tac [exI, conjI])));
```

```
Level 1
EX XA XB YA YB.
   XA Int XB = 0 &
   XA Un XB = X &
   YA Int YB = 0 & YA Un YB = Y & f `` XA = YA & g `` YB = XB

 1. ?XA Int ?XB1 = 0
 2. ?XA Un ?XB1 = X
 3. ?YA2 Int ?YB3 = 0
 4. ?YA2 Un ?YB3 = Y
 5. f `` ?XA = ?YA2
 6. g `` ?YB3 = ?XB1
```

The next command solves five of these subgoals by repeatedly applying facts such as $A \cap (B - A) = \emptyset$. Observe how the unknowns are instantiated; only `?XA` is left.

```
by (REPEAT
    (FIRSTGOAL (resolve_tac [Diff_disjoint, Diff_partition, refl])));
```

```
Level 2
EX XA XB YA YB.
   XA Int XB = 0 &
   XA Un XB = X &
   YA Int YB = 0 & YA Un YB = Y & f `` XA = YA & g `` YB = XB

 1. ?XA <= X
 2. f `` ?XA <= Y
 3. g `` (Y - f `` ?XA) = X - ?XA
```

We apply the result proved in the previous section to subgoal 3. This instantiates the last unknown to `lfp`$(\cdots)$:

```
by (resolve_tac [Banach_last_equation] 3);
```

```
Level 3
EX XA XB YA YB.
   XA Int XB = 0 &
   XA Un XB = X &
   YA Int YB = 0 & YA Un YB = Y & f `` XA = YA & g `` YB = XB
 1. lfp(X,%W. X - g `` (Y - f `` W)) <= X
 2. f `` lfp(X,%W. X - g `` (Y - f `` W)) <= Y
 3. g : Y -> X
```

The remaining subgoals are verified by appealing to lemmas and the premises.

```
by (REPEAT (resolve_tac (prems@[fun_is_rel, image_subset,
                             lfp_subset, decomp_bnd_mono]) 1));
 Level 4
 EX XA XB YA YB.
    XA Int XB = 0 &
    XA Un XB = X &
    YA Int YB = 0 & YA Un YB = Y & f `` XA = YA & g `` YB = XB
 No subgoals!
```

2.7.3. *The Schröder-Bernstein Theorem*
The Schröder-Bernstein Theorem is stated as

$$\frac{f \in \texttt{inj}(X,Y) \quad g \in \texttt{inj}(Y,X)}{\exists h \,.\, h \in \texttt{bij}(X,Y)}$$

The standard Isabelle proof consists of an appeal to Banach's Theorem and a call to the classical reasoner (`fast_tac`). Banach's Theorem introduces an existentially quantified assumption. The classical reasoner strips those quantifiers, adding new bound variables $X_A$, $X_B$, $Y_A$ and $Y_B$, to the context; then it strips the existential quantifier from the goal, yielding an unknown; finally it instantiates that unknown with a suitable bijection.

The form of the bijection is forced by the following three lemmas, which come from a previously developed library of permutations:

$$\frac{f \in \texttt{bij}(A,B) \quad g \in \texttt{bij}(C,D) \quad A \cap C = \emptyset \quad B \cap D = \emptyset}{f \cup g \in \texttt{bij}(A \cup C, B \cup D)} \; (\texttt{bij_disjoint_Un})$$

$$\frac{f \in \texttt{bij}(A,B)}{f^{-1} \in \texttt{bij}(B,A)} \; (\texttt{bij_converse_bij}) \qquad \frac{f \in \texttt{bij}(A,B) \quad C \subseteq A}{f \restriction C \in \texttt{bij}(C, f``C)} \; (\texttt{restrict_bij})$$

To demonstrate how the bijection is instantiated, let us state the theorem using an unknown rather than a existential quantifier. This proof requires supplying as premises the conclusions of Banach's Theorem *without* their existential quantifiers:

```
val prems = goal Cardinal.thy
    "[| f : inj(X,Y)  ;   g : inj(Y,X) ;         \
\       XA Int XB = 0 ;   XA Un XB = X ;         \
\       YA Int YB = 0 ;   YA Un YB = Y ;         \
\       f``XA = YA    ;   g``YB = XB   |] ==>   ?h: bij(X,Y)";
```

```
Level 0
?h : bij(X,Y)
 1. ?h : bij(X,Y)
```

The first step inserts the premises into subgoal 1 and performs all possible substitutions, such as $Y$ to $Y_A \cup Y_B$ and $Y_A$ to $f``X_A$.

```
by (cut_facts_tac prems 1  THEN
    REPEAT (hyp_subst_tac 1)  THEN  flexflex_tac);
```

```
Level 1
?h69 : bij(X,Y)
 1. [| f : inj(XA Un g `` YB,f `` XA Un YB);
       g : inj(f `` XA Un YB,XA Un g `` YB); XA Int g `` YB = 0;
       f `` XA Int YB = 0 |] ==>
    ?h69 : bij(XA Un g `` YB,f `` XA Un YB)
```

The second step applies `bij_disjoint_Un`, instantiating the bijection to consist of some union.

```
by (resolve_tac [bij_disjoint_Un] 1 THEN REPEAT (assume_tac 3));
```

```
Level 2
?f70 Un ?g70 : bij(X,Y)
 1. [| f : inj(XA Un g `` YB,f `` XA Un YB);
       g : inj(f `` XA Un YB,XA Un g `` YB); XA Int g `` YB = 0;
       f `` XA Int YB = 0 |] ==>
    ?f70 : bij(XA,f `` XA)

 2. [| f : inj(XA Un g `` YB,f `` XA Un YB);
       g : inj(f `` XA Un YB,XA Un g `` YB); XA Int g `` YB = 0;
       f `` XA Int YB = 0 |] ==>
    ?g70 : bij(g `` YB,YB)
```

The third step applies `bij_converse_bij` to subgoal 2, instantiating the bijection with a `converse` term. This rule should only be used in the last resort, since it can be repeated indefinitely.

```
by (resolve_tac [bij_converse_bij] 2);
```

```
Level 3
?f70 Un converse(?f71) : bij(X,Y)
 1. [| f : inj(XA Un g `` YB,f `` XA Un YB);
       g : inj(f `` XA Un YB,XA Un g `` YB); XA Int g `` YB = 0;
       f `` XA Int YB = 0 |] ==>
    ?f70 : bij(XA,f `` XA)
```

```
  2. [| f : inj(XA Un g `` YB,f `` XA Un YB);
        g : inj(f `` XA Un YB,XA Un g `` YB); XA Int g `` YB = 0;
        f `` XA Int YB = 0 |] ==>
     ?f71 : bij(YB,g `` YB)
```

The fourth step applies `restrict_bij`, instantiating the bijection with restrictions. We obtain $(f \upharpoonright X_A) \cup (g \upharpoonright Y_B)^{-1}$.

```
by (REPEAT (FIRSTGOAL (eresolve_tac [restrict_bij])));
 Level 4
 restrict(f,XA) Un converse(restrict(g,YB)) : bij(X,Y)
  1. [| g : inj(f `` XA Un YB,XA Un g `` YB); XA Int g `` YB = 0;
        f `` XA Int YB = 0 |] ==>
     XA <= XA Un g `` YB
  2. [| f : inj(XA Un g `` YB,f `` XA Un YB); XA Int g `` YB = 0;
        f `` XA Int YB = 0 |] ==>
     YB <= f `` XA Un YB
```

Finally we appeal to some obvious facts.

```
by (REPEAT (resolve_tac [Un_upper1,Un_upper2] 1));
 Level 5
 restrict(f,XA) Un converse(restrict(g,YB)) : bij(X,Y)
 No subgoals!
```

The total execution time to prove the Banach and Schröder-Bernstein Theorems is about three seconds.[3]

The Schröder-Bernstein Theorem is a long-standing challenge problem; both Bledsoe [3, page 31] and McDonald and Suppes [14, page 338] mention it. The Isabelle proof cannot claim to be automatic — it draws upon a body of lemmas — but it is short and comprehensible. It demonstrates the power of instantiating unknowns incrementally.

This mechanized theory of least fixedpoints allows formal reasoning about any inductively-defined subset of an existing set. Before we can use the theory to specify recursive data structures, we need some means of constructing large sets. Since large sets could be defined by transfinite recursion, we now consider the general question of recursive functions in set theory.

## 3. Recursive Functions

A relation $\prec$ is *well-founded* if it admits no infinite decreasing chains

$$\cdots \prec x_n \prec \cdots \prec x_2 \prec x_1.$$

Well-founded relations are a general means of justifying recursive definitions and proving termination. They have played a key role in the Boyer/Moore Theorem

Prover since its early days [4]. Manna and Waldinger's work on deductive program synthesis [12] illustrates the power of well-founded relations; they justify the termination of a unification algorithm using a relation that takes into account the size of a term and the number of free variables it contains.

The rise of type theory [6, 9, 13] has brought a new treatment of recursion. Instead of a single recursion operator justified by well-founded relations, each recursive type comes equipped with a structural recursion operator. For the natural numbers, structural recursion admits calls such as $\texttt{double}(n+1) = \texttt{double}(n)+2$; for lists, it admits calls such as $\texttt{rev}(\texttt{Cons}(x,l)) = \texttt{rev}(l)@[x]$.

These recursion operators are powerful — unlike computation theory's primitive recursion, they can express Ackermann's function — but they are sometimes inconvenient. They can only express recursive calls involving an immediate component of the argument. This excludes functions that divide by repeated subtraction or that sort by recursively sorting shorter lists. Coding such functions using structural recursion requires ingenuity; consider Smith's treatment of Quicksort [26].

Nordström [19] and I [21] have attempted to re-introduce well-founded relations to type theory, with limited success. In ZF set theory, well-founded relations reclaim their role as the foundation of induction and recursion. They can express difficult termination arguments, such as for unification and Quicksort; they include structural recursion as a special case.

Suppose we have defined the operator `list` such that $\texttt{list}(A)$ is the set of all lists of the form

$$\texttt{Cons}(x_1, \texttt{Cons}(x_2, \ldots, \texttt{Cons}(x_n, \texttt{Nil})\ldots)) \qquad x_1, x_2, \ldots, x_n \in A$$

We could then define the substructure relation $\texttt{is_tail}(A)$ to consist of all pairs $\langle l, \texttt{Cons}(x,l)\rangle$ for $x \in A$ and $l \in \texttt{list}(A)$, since $l$ is the tail of $\texttt{Cons}(x,l)$. Proving that $\texttt{is_tail}(A)$ is well-founded justifies structural recursion on lists.

But this approach can be streamlined. The well-foundedness of lists, trees and many similar data structures follows from the well-foundedness of ordered pairing, which follows from the Foundation Axiom of ZF set theory.[4] This spares us the effort of defining relations such as $\texttt{is_tail}(A)$. Moreover, recursive functions defined using $\texttt{is_tail}(A)$ have a needless dependence upon $A$; exploiting the Foundation Axiom eliminates this extra argument.

Achieving these aims requires considerable effort. Several highly technical set-theoretic constructions are defined in succession:

- A *well-founded recursion* operator, called `wfrec`, is defined and proved to satisfy a general recursion equation.
- The *ordinals* are constructed. Transfinite recursion is an instance of well-founded recursion.
- The *natural numbers* are constructed. Natural numbers are ordinals and inherit many of their properties from the ordinals. Primitive recursion on the natural numbers is an instance of transfinite recursion.

– The *rank* operation associates a unique ordinal with every set; it serves as an absolute measure of a set's depth. To define this operation, transfinite recursion is generalized to a form known as $\in$-recursion (`transrec` in Isabelle ZF). The construction involves the natural numbers.

– The *cumulative hierarchy* of ZF set theory is finally introduced, by transfinite recursion. As a special case, it includes a small 'universe' for use with `lfp` in defining recursive data structures.

– The general recursion operator `Vrec` justifies functions that make recursive calls on arguments of lesser rank.

## 3.1. Well-Founded Recursion

The ZF derivation of well-founded recursion is based on one by Tobias Nipkow in higher-order logic. It is much shorter than any other derivation that I have seen, including several of my own. It is still complex, more so than a glance at Suppes [27, pages 197–8] might suggest. Space permits only a discussion of the definitions and key theorems.

### 3.1.1. *Definitions*

First, we must define 'well-founded relation.' Infinite descending chains are difficult to formalize; a simpler criterion is that each non-empty set contains a minimal element. The definition echoes the Axiom of Foundation [22, §4].

$$\texttt{wf}(r) \;\equiv\; \forall Z \,.\, Z = \emptyset \lor (\exists_{x \in Z} \,.\, \forall y \,.\, \langle y, x\rangle \in r \to y \notin Z)$$

From this, it is fairly easy to derive well-founded induction:

$$\frac{\texttt{wf}(r) \qquad \begin{array}{c}[\forall y \,.\, \langle y, x\rangle \in r \to \psi(y)]_x \\ \vdots \\ \psi(x)\end{array}}{\psi(a)}$$

Proof: put $\{z \in \texttt{domain}(r) \cup \{a\} \,.\, \neg\psi(z)\}$ for $Z$ in $\texttt{wf}(r)$. If $Z = \emptyset$ then $\psi(a)$ follows immediately. If $Z$ is nonempty then we obtain an $x$ such that $\neg\psi(x)$ and (by the definition of `domain`) $\forall y \,.\, \langle y, x\rangle \in r \to \psi(y)$, but the latter implies $\psi(x)$. The Isabelle proof is only seven lines.

Well-founded recursion, on the other hand, is difficult even to formalize. If $f$ is recursive over the well-founded relation $r$ then $f`x$ may depend upon $x$ and, for $\langle y, x\rangle \in r$, upon $f`y$. Since $f$ need not be computable, $f`x$ may depend upon infinitely many values of $f`y$. The inverse image $r^{-1}``\{x\}$ is the set of all $y$ such that $\langle y, x\rangle \in r$: the set of all *r-predecessors* of $x$. Formally, $f$ is recursive over $r$ if it satisfies the equation

$$f`x \;=\; H(x, f \restriction (r^{-1}``\{x\})) \tag{3}$$

for all $x$. The binary operation $H$ is the body of $f$. Restricting $f$ to $r^{-1}``\{x\}$ ensures that the argument in each recursive call is $r$-smaller than $x$.

Justifying well-founded recursion requires proving, for all $r$ and $H$, that the corresponding recursive function exists. It is constructed in stages by well-founded induction. Call $f$ a *restricted recursive function for $x$* if it satisfies equation (3) for all $y$ such that $\langle y, x\rangle \in r$. For a fixed $x$, we assume there exist restricted recursive functions for all the $r$-predecessors of $x$, and construct from them a restricted recursive function for $x$. We must also show that the restricted recursive functions agree where their domains overlap; this ensures that the functions are unique.

Nipkow's formalization of the construction makes several key simplifications. Since the transitive closure $r^+$ of a well-founded relation $r$ is well-founded, he restricts the construction to transitive relations; otherwise it would have to use $r$ in some places and $r^+$ in others, leading to complications. Second, he formalizes '$f$ is a restricted recursive function for $a$' by a neat equation:

$$\texttt{is_recfun}(r, a, H, f) \;\equiv\; (f = \lambda x \in r^{-1}``\{a\} \,.\, H(x, f \restriction (r^{-1}``\{x\})))$$

Traditional proofs define the full recursive function as the union of all restricted recursive functions. This involves tiresome reasoning about sets of ordered pairs. Nipkow instead uses descriptions:

$$\begin{aligned} \texttt{the_recfun}(r, a, H) \;&\equiv\; \iota f \,.\, \texttt{is_recfun}(r, a, H, f) \\ \texttt{wftrec}(r, a, H) \;&\equiv\; H(a, \texttt{the_recfun}(r, a, H)) \end{aligned}$$

Here $\texttt{the_recfun}(r, a, H)$ denotes the (unique) restricted recursive function for $a$. Finally, wftrec gives access to the full recursive function; $\texttt{wftrec}(r, a, H)$ yields the result for the argument $a$.

3.1.2. *Lemmas*

Here are the key lemmas. Assume $\texttt{wf}(r)$ and $\texttt{trans}(r)$ below, where $\texttt{trans}(r)$ expresses that $r$ is transitive.

Two restricted recursive functions $f$ and $g$ agree over the intersection of their domains — by well-founded induction on $x$:

$$\frac{\texttt{is_recfun}(r, a, H, f) \quad \texttt{is_recfun}(r, b, H, g)}{\langle x, a\rangle \in r \land \langle x, b\rangle \in r \to f`x = g`x}$$

In consequence, the restricted recursive function at $a$ is unique:

$$\frac{\texttt{is_recfun}(r, a, H, f) \quad \texttt{is_recfun}(r, a, H, g)}{f = g}$$

Another consequence justifies our calling such functions 'restricted,' since they are literally restrictions of larger functions:

$$\frac{\texttt{is_recfun}(r, a, H, f) \quad \texttt{is_recfun}(r, b, H, g) \quad \langle b, a\rangle \in r}{f \restriction (r^{-1}``\{b\}) = g}$$

Using well-founded induction again, we prove the key theorem. Restricted recursive functions exist for all $a$:

$$\texttt{is_recfun}(r, a, H, \texttt{the_recfun}(r, a, H))$$

It is now straightforward to prove that wftrec unfolds as desired for well-founded recursion:

$$\texttt{wftrec}(r, a, H) = H(a, \lambda x \in r^{-1}``\{a\} \,.\, \texttt{wftrec}(r, x, H))$$

The abstraction over $r^{-1}``\{a\}$ is essentially the same as restriction.

3.1.3. *The Recursion Equation*

It only remains to remove the assumption $\texttt{trans}(r)$. Because the transitive closure of a well-founded relation is well-founded, we can immediately replace $r$ by $r^+$ in the recursion equation for wftrec. But this leads to strange complications later, involving transfinite recursion. I find it better to remove transitive closure from the recursion equation, even at the cost of weakening it.[5] The operator wfrec applies wftrec with the transitive closure of $r$, but restricts recursive calls to immediate $r$-predecessors:

$$\texttt{wfrec}(r, a, H) \equiv \texttt{wftrec}(r^+, a, \lambda x f \,.\, H(x, f \restriction (r^{-1}``\{x\})))$$

Assuming $\texttt{wf}(r)$ but not $\texttt{trans}(r)$, we can show the equation for wfrec:

$$\texttt{wfrec}(r, a, H) = H(a, \lambda x \in r^{-1}``\{a\} \,.\, \texttt{wfrec}(r, x, H))$$

All recursive functions in Isabelle's ZF set theory are ultimately defined in terms of wfrec.

## 3.2. Ordinals

My treatment of recursion requires a few properties of the set-theoretic ordinals. The development follows standard texts [27] and requires little further discussion. By convention, the Greek letters $\alpha$, $\beta$ and $\gamma$ range over ordinals.

A set $A$ is *transitive* if it is downwards closed under the membership relation: $y \in x \in A$ implies $y \in A$. An *ordinal* is a transitive set whose elements are also transitive. The elements of an ordinal are therefore ordinals also. The finite ordinals are the natural numbers; the set of natural numbers is itself an ordinal, called $\omega$. *Transfinite* ordinals are those greater than $\omega$; they serve many purposes in set theory and are the key to the recursion principles discussed below.

The Isabelle definitions are routine. The predicates Transset and Ord define transitive sets and ordinals, while $<$ is the less-than relation on ordinals:

$$\begin{aligned}
\texttt{Memrel}(A) &\equiv \{z \in A \times A \,.\, \exists x\, y \,.\, z = \langle x, y\rangle \wedge x \in y\} \\
\texttt{Transset}(i) &\equiv \forall_{x \in i} \,.\, x \subseteq i \\
\texttt{Ord}(i) &\equiv \texttt{Transset}(i) \wedge (\forall_{x \in i} \,.\, \texttt{Transset}(x)) \\
i < j &\equiv i \in j \wedge \texttt{Ord}(j)
\end{aligned}$$

The set $\texttt{Memrel}(A)$ internalizes the membership relation on $A$ as a subset of $A \times A$. If $A$ is transitive then $\texttt{Memrel}(A)$ internalizes the membership relation everywhere below $A$. For then

$$x_1 \in x_2 \in \cdots \in x_n \in A$$

implies that $x_1$, $x_2$, ..., $x_n$ are all elements of $A$; we have $\langle x_k, x_{k+1}\rangle \in \texttt{Memrel}(A)$ for $0 < k < n$.

A common use of $\texttt{wfrec}$ has the form $\texttt{wfrec}(\texttt{Memrel}(A), x, H)$, where $A$ is a transitive set and $x \in A$. The recursion equation for $\texttt{wfrec}(\texttt{Memrel}(A), x, H)$ supplies $\texttt{Memrel}(A)$ as the well-founded relation in the recursive calls. We must use $\texttt{Memrel}(A)$ because well-founded induction and recursion take their well-founded relation as a set, not as a binary predicate such as $\in$.

Using the Foundation Axiom, it is straightforward to show that $\texttt{Memrel}(A)$ is well-founded. This fact, together with the transitivity of ordinals, yields transfinite induction:

$$\frac{\texttt{Ord}(\alpha) \qquad \begin{matrix}[\texttt{Ord}(\beta) \quad \forall_{\gamma\in\beta} \,.\, \psi(\gamma)]_\beta \\ \vdots \\ \psi(\beta)\end{matrix}}{\psi(\alpha)}$$

Many properties of the ordinals are established by transfinite induction. For example, the ordinals are linearly ordered:

$$\frac{\texttt{Ord}(\alpha) \quad \texttt{Ord}(\beta)}{\alpha < \beta \vee \alpha = \beta \vee \beta < \alpha}$$

The *successor* of $x$, written $\texttt{succ}(x)$, is traditionally defined by $\texttt{succ}(x) \equiv \{x\} \cup x$. The Isabelle theory makes an equivalent definition using $\texttt{cons}$:

$$\texttt{succ}(x) \;\equiv\; \texttt{cons}(x, x)$$

Successors have two key properties:

$$\frac{\texttt{succ}(x) = \texttt{succ}(y)}{x = y} \qquad \texttt{succ}(x) \neq 0$$

Proving that $\texttt{succ}$ is injective seems to require the Axiom of Foundation. Proving $\texttt{succ}(x) \neq 0$ is trivial because zero is the empty set; let us write the empty set as 0 instead of $\emptyset$ when it serves as zero.

The smallest ordinal is zero. The ordinals are closed under the successor operation. The union of any family of ordinals is itself an ordinal, which happens to be their least upper bound:

$$\texttt{Ord}(0) \qquad \frac{\texttt{Ord}(\alpha)}{\texttt{Ord}(\texttt{succ}(\alpha))} \qquad \frac{\begin{matrix}[x \in A]_x \\ \vdots \\ \texttt{Ord}(\beta(x))\end{matrix}}{\texttt{Ord}(\bigcup_{x\in A} .\beta(x))}$$

By the first two rules above, every natural number is an ordinal. By the third, so is the set of natural numbers. This ordinal is traditionally called $\omega$; the following section defines it as the set `nat`.

Transfinite recursion can be expressed using `wfrec` and `Memrel`; see `nat_rec` below. Later (§3.4) we shall define a more general form of transfinite recursion, called $\in$-recursion.

## 3.3. The Natural Numbers

The natural numbers are a recursive data type, but they must be defined now (a bit prematurely) in order to complete the development of the recursion principles. The operator `nat_case` provides case analysis on whether a natural number has the form 0 or $\texttt{succ}(k)$, while `nat_rec` is a structural recursion operator similar to those in Martin-Löf's Type Theory [13].

$$\begin{aligned} \texttt{nat} &\equiv \texttt{lfp}(\texttt{Inf},\ \lambda X\,.\,\{0\} \cup \{\texttt{succ}(i)\,.\,i \in X\}) \\ \texttt{nat_case}(a,b,k) &\equiv \iota y\,.\,(k = 0 \wedge y = a) \vee (\exists i\,.\,k = \texttt{succ}(i) \wedge y = b(i)) \\ \texttt{nat_rec}(a,b,k) &\equiv \texttt{wfrec}(\texttt{Memrel}(\texttt{nat}), k,\ \lambda n f\,.\,\texttt{nat_case}(a,\ \lambda m\,.\,b(m, f`m), n)) \end{aligned}$$

Each definition is discussed below. They demonstrate the Knaster-Tarski Theorem, descriptions, and well-founded recursion.

### 3.3.1. *Properties of* `nat`

The mapping supplied to `lfp`, which takes $X$ to $\{0\}\cup\{\texttt{succ}(i).i \in X\}$, is obviously monotonic. The Axiom of Infinity supplies the constant `Inf` for the bounding set:[6]

$$(0 \in \texttt{Inf}) \wedge (\forall_{y\in\texttt{Inf}}\,.\,\texttt{succ}(y) \in \texttt{Inf})$$

The Axiom gives us a set containing zero and closed under the successor operation; the least such set contains nothing but the natural numbers.

The Knaster-Tarski Theorem yields

$$\texttt{nat} = \{0\} \cup \{\texttt{succ}(i)\,.\,i \in \texttt{nat}\}$$

and we immediately obtain the introduction rules

$$0 \in \texttt{nat} \qquad \frac{n \in \texttt{nat}}{\texttt{succ}(n) \in \texttt{nat}}$$

By instantiating the general induction rule of `lfp`, we obtain mathematical induction (recall our discussion in §2.3 above):

$$\frac{n \in \texttt{nat} \quad \psi(0) \quad \begin{array}{c}[x \in \texttt{nat} \quad \psi(x)]_x \\ \vdots \\ \psi(\texttt{succ}(x))\end{array}}{\psi(x)}$$

3.3.2. *Properties of* nat_case

The definition of nat_case contains a typical definite description. Given theorems stating $\mathtt{succ}(m) \neq 0$ and $\mathtt{succ}(m) = \mathtt{succ}(n) \to m = n$, Isabelle's fast_tac automatically proves the key equations:

$$\mathtt{nat_case}(a, b, 0) = a \qquad \mathtt{nat_case}(a, b, \mathtt{succ}(m)) = b(m)$$

3.3.3. *Properties of* nat_rec

Because nat is an ordinal, it is a transitive set. Well-founded recursion on Memrel(nat), which denotes the less-than relation on the natural numbers, can express primitive recursion. Unfolding the recursion equation for wfrec yields

$$\mathtt{nat_rec}(a, b, n) = \mathtt{nat_case}(a,\ \lambda m\,.\,b(m, f`m),\ n)$$

where $f \equiv \lambda x \in \mathtt{Memrel}(\mathtt{nat})^{-1}``\{n\}.\mathtt{nat_rec}(a, b, x)$. We may derive the equations

$$\mathtt{nat_rec}(a, b, 0) = a \qquad \frac{m \in \mathtt{nat}}{\mathtt{nat_rec}(a, b, \mathtt{succ}(m)) = b(m, \mathtt{nat_rec}(a, b, m))}$$

The first equation is trivial, by the similar one for nat_case. Assuming $m \in \mathtt{nat}$, the second equation follows by $\beta$-conversion. This requires showing

$$m \in \mathtt{Memrel}(\mathtt{nat})^{-1}``\{\mathtt{succ}(m)\},$$

which reduces to

$$\langle m, \mathtt{succ}(m)\rangle \in \mathtt{Memrel}(\mathtt{nat}),$$

and finally to the trivial

$$m \in \mathtt{succ}(m) \qquad m \in \mathtt{nat} \qquad \mathtt{succ}(m) \in \mathtt{nat}.$$

The Isabelle proofs of these rules are straightforward. Recursive definitions of lists and trees will follow the pattern established above. But first, we must define transfinite recursion in order to construct large sets.

## 3.4. The Rank Function

Many of the ZF axioms assert the existence of sets, but all sets can be generated in a uniform manner. Each stage of the construction is labelled by an ordinal $\alpha$; the set of all sets generated by stage $\alpha$ is called $V_\alpha$. Each stage simply gathers up the powersets of all the previous stages. Define

$$V_\alpha \;=\; \bigcup_{\beta\in\alpha} \wp(V_\beta)$$

by transfinite recursion on the ordinals. In particular we have $V_0 = \emptyset$ and $V_{\mathtt{succ}(\alpha)} = \wp(V_\alpha)$. See Devlin [8, pages 42–48] for philosophy and discussion.

We can define the ordinal $\mathtt{rank}(a)$, for all sets $a$, such that $a \subseteq V_{\mathtt{rank}(a)}$. This attaches an ordinal to each and every set, indicating the stage of its creation. When seeking a large 'bounding set' for use with the `lfp` operator, we can restrict our attention to sets of the form $V_\alpha$, since every set is contained in some $V_\alpha$.

Taken together, the $V_\alpha$ are called the *cumulative hierarchy*. They are fundamental to the intuition of set theory, since they impart a structure to the universe of sets. Their role here is more mundane. We need $\mathtt{rank}(a)$ and $V_\alpha$ to apply `lfp` and to justify structural recursion. The following section will formalize the definition of $V_\alpha$.

3.4.1. *Informal Definition of* `rank`

The usual definition of `rank` requires $\in$-recursion:

$$\mathtt{rank}(a) = \bigcup_{x\in a} \mathtt{succ}(\mathtt{rank}(x))$$

The recursion resembles that of $V_\alpha$, except that it is not restricted to the ordinals. Recursion over the ordinals is straightforward because each ordinal is transitive (recall the discussion in §3.2). To justify $\in$-recursion, we define an operation `eclose`, such that $\mathtt{eclose}(a)$ extends $a$ to be transitive. Let $\bigcup^n(X)$ denote the $n$-fold union of $X$, with $\bigcup^0(X) = X$ and $\bigcup^{\mathtt{succ}(m)}(X) = \bigcup(\bigcup^m(X))$. Then put

$$\mathtt{eclose}(a) = \bigcup_{n\in\mathtt{nat}} \bigcup^n(a)$$

and supply $\mathtt{Memrel}(\mathtt{eclose}(\{a\}))$ as the well-founded relation for recursion on $a$.

3.4.2. *The Formal Definitions*

Here are the Isabelle definitions of `eclose`, `transrec` (which performs $\in$-recursion) and `rank`:

$$\begin{aligned}
\mathtt{eclose}(a) &\equiv \bigcup_{n\in\mathtt{nat}} \mathtt{nat_rec}(a,\ \lambda m\, r\, .\bigcup(r),\ n)\\
\mathtt{transrec}(a, H) &\equiv \mathtt{wfrec}(\mathtt{Memrel}(\mathtt{eclose}(a)), a, H)\\
\mathtt{rank}(a) &\equiv \mathtt{transrec}(a,\ \lambda x f\, .\bigcup_{y\in x} \mathtt{succ}(f`y))
\end{aligned}$$

3.4.3. *The Main Theorems*

Many results are proved about `eclose`; the most important perhaps is that $\mathtt{eclose}(a)$ is the smallest transitive set containing $a$. Now $\mathtt{Memrel}(\mathtt{eclose}(\{a\}))$ contains enough of the membership relation to include every chain $x_1 \in \cdots \in x_n \in a$ descending from $a$. As an instance of well-founded induction, we obtain

$\in$-induction:

$$\frac{\begin{array}{c}[\forall_{y\in x}\,.\,\psi(y)]_x\\ \vdots\\ \psi(x)\end{array}}{\psi(a)}$$

Now $\in$-recursion follows similarly, but there is another technical hurdle. In $\texttt{transrec}(a,H)$, the well-founded relation supplied to `wfrec` depends upon $a$; we must show that the result of `wfrec` does not depend upon the field of the relation $\texttt{Memrel}(\texttt{eclose}(\cdots))$, if it is big enough. Specifically, we must show

$$\frac{k\in i}{\texttt{wfrec}(\texttt{Memrel}(\texttt{eclose}(\{i\})),k,H)=\texttt{wfrec}(\texttt{Memrel}(\texttt{eclose}(\{k\})),k,H)}$$

in order to derive the recursion equation

$$\texttt{transrec}(a,H)=H(a,\lambda_{x\in a}\,.\,\texttt{transrec}(x,H)).$$

Combining this with the definition of `rank` yields

$$\texttt{rank}(a)=\bigcup_{y\in a}\texttt{succ}(\texttt{rank}(y)).$$

Trivial transfinite inductions prove $\texttt{Ord}(\texttt{rank}(a))$ and $\texttt{rank}(\alpha)=\alpha$ for ordinals $\alpha$.

We may use `rank` to measure the depth of a set. The following facts will justify recursive function definitions over lists and trees by proving that the recursion is well-founded:

$$\frac{a\in b}{\texttt{rank}(a)<\texttt{rank}(b)}\qquad \texttt{rank}(a)<\texttt{rank}(\langle a,b\rangle)\qquad \texttt{rank}(b)<\texttt{rank}(\langle a,b\rangle)$$

Let us prove the last of these from the first. Recall from Part I [22, §7.3] the definition of ordered pairs, $\langle a,b\rangle\equiv\{\{a\},\{a,b\}\}$. From $b\in\{a,b\}$ we obtain $\texttt{rank}(b)<\texttt{rank}(\{a,b\})$. From $\{a,b\}\in\langle a,b\rangle$ we obtain $\texttt{rank}(\{a,b\})<\texttt{rank}(\langle a,b\rangle)$. Now $<$ is transitive, yielding $\texttt{rank}(b)<\texttt{rank}(\langle a,b\rangle)$.

We need $\in$-recursion only to define `rank`, since this operator can reduce every other instance of $\in$-recursion to transfinite recursion on the ordinals. We shall use `transrec` immediately below and `rank` in the subsequent section.

### 3.5. The Cumulative Hierarchy

We can now formalize the definition $V_\alpha=\bigcup_{\beta\in\alpha}\wp(V_\beta)$, which was discussed above. A useful generalization is to construct the cumulative hierarchy starting from a given set $A$:

$$V[A]_\alpha\;=\;A\cup\bigcup_{\beta\in\alpha}\wp(V[A]_\beta)\tag{4}$$

Later, $V[A]_\omega$ will serve as a 'universe' for defining recursive data structures; it contains all finite lists and trees built over $A$. The Isabelle definitions include

$$V[A]_\alpha \equiv \texttt{transrec}(\alpha,\ \lambda\alpha f\,.\, A \cup \bigcup_{\beta\in\alpha} \wp(V_\beta))$$
$$V_\alpha \equiv V[\emptyset]_\alpha$$

3.5.1. *Closure Properties of* $V[A]_\alpha$

The Isabelle ZF theory proves several dozen facts involving $V[A]_\alpha$. Because its definition uses $\in$-recursion, $V[A]_x$ is meaningful for every set $x$. But the most important properties concern $V[A]_\alpha$ where $\alpha$ is an ordinal. Many are proved by transfinite induction on $\alpha$.

To justify the term 'cumulative hierarchy,' note that $V[A]_x$ is monotonic in both $A$ and $x$:

$$\frac{A \subseteq B \quad x \subseteq y}{V[A]_x \subseteq V[B]_y}$$

For ordinals we obtain $V[A]_\alpha \subseteq V[A]_{\texttt{succ}(\alpha)}$ as a corollary.

The cumulative hierarchy satisfies several closure properties. Here are three elementary ones:

$$x \subseteq V[A]_x \qquad A \subseteq V[A]_x \qquad \frac{a \subseteq V[A]_\alpha}{a \in V[A]_{\texttt{succ}(\alpha)}}$$

By the third property, increasing the ordinal generates finite sets:

$$\frac{a_1 \in V[A]_\alpha \quad \cdots \quad a_n \in V[A]_\alpha}{\{a_1, \ldots, a_n\} \in V[A]_{\texttt{succ}(\alpha)}}$$

Since $\langle a, b\rangle \equiv \{\{a\}, \{a, b\}\}$, increasing the ordinal twice generates ordered pairs:

$$\frac{a \in V[A]_\alpha \quad b \in V[A]_\alpha}{\langle a, b\rangle \in V[A]_{\texttt{succ}(\texttt{succ}(\alpha))}}$$

Now put $\alpha = \omega$, recalling that $\omega$ is just the set `nat` of all natural numbers. Let us prove that $V[A]_\omega$ is closed under products:

$$V[A]_\omega \times V[A]_\omega \subseteq V[A]_\omega$$

Suppose we have $a$, $b \in V[A]_\omega$. By equation (4), there exist $i$, $j \in \texttt{nat}$ such that $a \in V[A]_i$ and $b \in V[A]_j$. Let $k$ be the greater of $i$ and $j$; then $a$, $b \in V[A]_k$. Since $\langle a, b\rangle \in V[A]_{\texttt{succ}(\texttt{succ}(k))}$ and $\texttt{succ}(\texttt{succ}(k)) \in \texttt{nat}$, we conclude $\langle a, b\rangle \in V[A]_\omega$.

By a similar argument, every finite subset of $V[A]_\omega$ is an element of $V[A]_\omega$. These ordered pairs and finite subsets are ultimately constructed from natural numbers and elements of $A$, since $V[A]_\omega$ contains `nat` and $A$ as subsets.

A *limit ordinal* is one that is non-zero and closed under the successor operation:

$$\texttt{Limit}(\alpha) \;\equiv\; \texttt{Ord}(\alpha) \wedge 0 < \alpha \wedge (\forall y \,.\, y < \alpha \rightarrow \texttt{succ}(y) < \alpha)$$

The smallest limit ordinal is $\omega$. The closure properties just discussed of $V[A]_\omega$ hold when $\omega$ is replaced by any limit ordinal. We shall use these closure properties in §4.2.

## 3.6. Recursion on a Set's Rank

Consider using recursion over lists formed by repeated pairing. The tail of the list $\langle x, l\rangle$ is $l$. Since $l$ is not a member of the set $\langle x, l\rangle$, we cannot use $\in$-recursion to justify a recursive call on $l$. But $l$ has smaller rank than $\langle x, l\rangle$; since ordinals are well-founded, this ensures that the recursion terminates.

The following recursion operator allows any recursive calls involving sets of lesser rank. It handles the list example above, as well as recursive calls for components of deep nests of pairs:

$$\begin{aligned}\texttt{Vrec}(a, H) \;\equiv\; & \texttt{transrec}(\texttt{rank}(a),\\ & \quad \lambda i\, g \,.\, \lambda z \in V_{\texttt{succ}(i)} \,.\, H(z, \lambda y \in V_i \,.\, g\text{‘}\texttt{rank}(y)\text{‘}y))\text{ ‘ } a\end{aligned}$$

This definition looks complex, but its formal properties are easy to derive. The rest of this section attempts to convey the underlying intuition.

### 3.6.1. *The Idea Behind* `Vrec`

To understand the definition of `Vrec`, consider a technique for defining general recursive functions over the natural numbers. The definition is reduced to one involving a primitive recursive functional. Suppose we wish to define a function $f$ satisfying the recursion

$$f\text{‘}x = H(x, f).$$

Suppose that, for all $x$ in the desired domain of $H$, the number $k(x)$ exceeds the depth of recursive calls required to compute $f\text{‘}x$. Define the family of functions $\hat{f}_n$ by primitive recursion over $n$:

$$\begin{aligned}\hat{f}_0 \;&\equiv\; \lambda_{x\in\texttt{nat}} \,.\, x\\ \hat{f}_{n+1} \;&\equiv\; \lambda_{x\in\texttt{nat}} \,.\, H(x, \hat{f}_n)\end{aligned}$$

Clearly, $\hat{f}_n$ behaves like $f$ if the depth of recursive calls is smaller than $n$; the definition of $\hat{f}_0$ is wholly immaterial, since it is never used. We can therefore define $f \equiv \lambda_{x\in\texttt{nat}} \,.\, \hat{f}_{k(x)}\text{‘}x$.

### 3.6.2. *The Workings of* `Vrec`

The definition of `Vrec` follows a similar idea. Using transfinite recursion, define a family of functions $\hat{f}_\alpha$ such that

$$\hat{f}_\alpha\text{‘}x \;=\; H(x, \lambda y \in V_{\texttt{rank}(x)} \,.\, \hat{f}_{\texttt{rank}(y)}\text{‘}y) \tag{5}$$

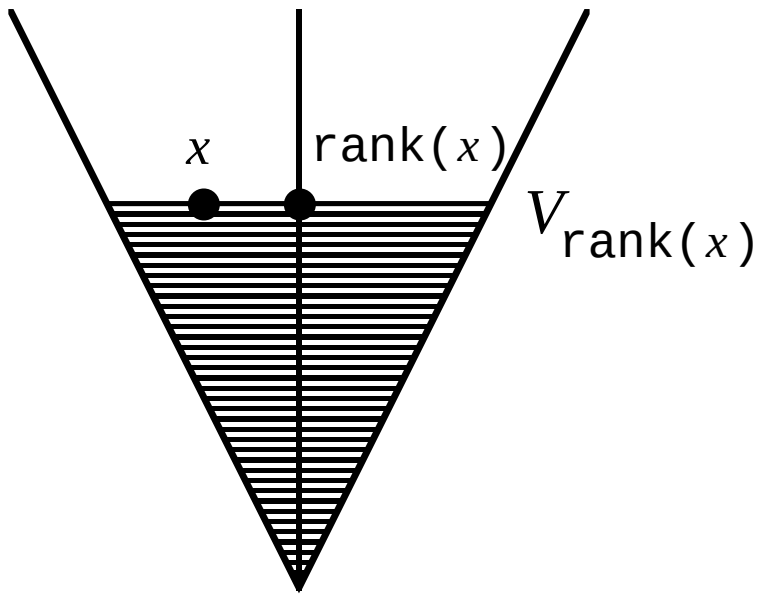

*Figure 2.* Domain for recursive calls in $\mathtt{Vrec}(x, H)$

for all $x$ in a sufficiently large set (which will depend upon $\alpha$), and define

$$\mathtt{Vrec}(x, H) \;\equiv\; \hat{f}_{\mathtt{rank}(x)} \,`\, x\,. \tag{6}$$

Here, $\mathtt{rank}(x)$ serves as an upper bound on the number of recursive calls required to compute $\mathtt{Vrec}(x, H)$. Combining equations (5) and (6) immediately yields the desired recursion:

$$\begin{aligned}\mathtt{Vrec}(x, H) &= H(x, \lambda y \in V_{\mathtt{rank}(x)} \,.\, \hat{f}_{\mathtt{rank}(y)}`y) \\ &= H(x, \lambda y \in V_{\mathtt{rank}(x)} \,.\, \mathtt{Vrec}(y, H))\end{aligned}$$

The key fact $y \in V_\alpha \leftrightarrow \mathtt{rank}(y) \in \alpha$ states that the set $V_\alpha$ consists of all sets whose rank is smaller than $\alpha$. For a given $x$, $\mathtt{Vrec}(x, H)$ may perform recursive calls for all $y$ of smaller rank than $x$ (see Figure 2). This general principle can express recursive functions for lists, trees and similar data structures based on ordered pairing.

We may formalize $\hat{f}_\alpha$ using $\mathtt{transrec}$:

$$\hat{f}_\alpha \;\equiv\; \mathtt{transrec}(\alpha,\; \lambda i\, g \,.\, \lambda z \in V_{\mathtt{succ}(i)} \,.\, H(z, \lambda y \in V_i \,.\, g`\mathtt{rank}(y)`y))$$

Unfolding $\mathtt{transrec}$ and simplifying yields equation (5), with $V_{\mathtt{succ}(\alpha)}$ as the 'sufficiently large set' mentioned above. Joining this definition with equation (6) yields the full definition of $\mathtt{Vrec}$.

The recursion equation for $\mathtt{Vrec}$ can be recast into a form that takes a definition in the premise:

$$\frac{\forall x \,.\, h(x) = \mathtt{Vrec}(x, H)}{h(a) = H(a,\, \lambda y \in V_{\mathtt{rank}(a)} \,.\, h(y))}$$

This expresses the recursion equation more neatly. The conclusion contains only one occurrence of $H$ instead of three, and $H$ is typically complex.

The following sections include worked examples using $\mathtt{Vrec}$ to express recursive functions.

## 4. Recursive Data Structures

This section presents ZF formalizations of lists and two different treatments of mutually recursive trees/forests. Before we can begin, two further tools are needed: disjoint sums and a 'universe' for solving recursion equations over sets.

### 4.1. Disjoint Sums

Let $1 \equiv \mathtt{succ}(0)$. Disjoint sums have a completely straightforward definition:

$$\begin{aligned} A + B &\equiv (\{0\} \times A) \cup (\{1\} \times B) \\ \mathtt{Inl}(a) &\equiv \langle 0, a \rangle \\ \mathtt{Inr}(b) &\equiv \langle 1, b \rangle \end{aligned}$$

We obtain the obvious introduction rules

$$\frac{a \in A}{\mathtt{Inl}(a) \in A + B} \qquad \frac{b \in B}{\mathtt{Inr}(b) \in A + B}$$

and other rules to state that `Inl` and `Inr` are injective and distinct. A `case` operation, defined by a description, satisfies two equations:

$$\mathtt{case}(c, d, \mathtt{Inl}(a)) = c(a) \qquad \mathtt{case}(c, d, \mathtt{Inr}(b)) = d(b)$$

This resembles the `when` operator of Martin-Löf's Type Theory [20].

### 4.2. A Universe

The term *universe* generally means the class of all sets, but here it refers to the set $\mathtt{univ}(A)$, which contains all finitely branching trees over $A$. The set is defined by

$$\mathtt{univ}(A) \equiv V[A]_\omega.$$

By the discussion of $V[A]_\omega$ in §3.5, we have

$$\mathtt{univ}(A) \times \mathtt{univ}(A) \subseteq \mathtt{univ}(A).$$

From the simpler facts $A \subseteq \mathtt{univ}(A)$ and $\mathtt{nat} \subseteq \mathtt{univ}(A)$, we obtain

$$\mathtt{univ}(A) + \mathtt{univ}(A) \subseteq \mathtt{univ}(A).$$

So $\mathtt{univ}(A)$ contains $A$ and the natural numbers, and is closed under disjoint sums and Cartesian products. We may use it with `lfp` to define lists and trees as least fixedpoints over $\mathtt{univ}(A)$, for a suitable set $A$.

Infinitely branching trees require larger universes. To construct them requires cardinality reasoning. Let $\kappa$ be an infinite cardinal. Writing the next larger cardinal as $\kappa^+$, a suitable universe for infinite branching up to $\kappa$ is $V[A]_{\kappa^+}$. I have recently formalized this approach in Isabelle's ZF set theory, proving the theorem $\kappa \to V[A]_{\kappa^+} \subseteq V[A]_{\kappa^+}$ and constructing an example with countable branching. The cardinality arguments appear to require the Axiom of Choice, and involve a large body of proofs. I plan to report on this work in a future paper.

## 4.3. Lists

Let $\texttt{list}(A)$ denote the set of all finite lists taking elements from $A$. Formally, $\texttt{list}(A)$ should satisfy the recursion $\texttt{list}(A) = \{\emptyset\} + A \times \texttt{list}(A)$. Since $\texttt{univ}(A)$ contains $\emptyset$ and is closed under $+$ and $\times$, it contains solutions of this equation. We simultaneously define the constructors Nil and Cons:[7]

$$\begin{aligned} \texttt{list}(A) &\equiv \texttt{lfp}(\texttt{univ}(A),\ \lambda X\,.\,\{\emptyset\} + A \times X) \\ \texttt{Nil} &\equiv \texttt{Inl}(\emptyset) \\ \texttt{Cons}(a,l) &\equiv \texttt{Inr}(\langle a,l\rangle) \end{aligned}$$

The mapping from $X$ to $\{\emptyset\} + A \times X$ is trivially monotonic by the rules shown in §2.4, and $\texttt{univ}(A)$ is closed under it. Therefore, the Knaster-Tarski Theorem yields $\texttt{list}(A) = \{\emptyset\} + A \times \texttt{list}(A)$ and we obtain the introduction rules:

$$\texttt{Nil} \in \texttt{list}(A) \qquad \frac{a \in A \quad l \in \texttt{list}(A)}{\texttt{Cons}(a,l) \in \texttt{list}(A)}$$

With equal ease, we derive structural induction for lists:

$$\frac{l \in \texttt{list}(A) \quad \psi(\texttt{Nil}) \quad \begin{matrix} [x \in A \quad y \in \texttt{list}(A) \quad \psi(y)]_{x,y} \\ \vdots \\ \psi(\texttt{Cons}(x,y)) \end{matrix}}{\psi(l)}$$

### 4.3.1. *Operating Upon Lists*

Again following Martin-Löf's Type Theory [13], we operate upon lists using case analysis and structural recursion. Here are their definitions in set theory:

$$\begin{aligned} \texttt{list_case}(c,h,l) &\equiv \texttt{case}(\lambda u\,.\,c, \texttt{split}(h), l) \\ \texttt{list_rec}(c,h,l) &\equiv \texttt{Vrec}(l,\ \lambda l\,g\,.\,\texttt{list_case}(c,\ \lambda x\,y\,.\,h(x,y,g`y),\ l)) \end{aligned}$$

Recall from Part I [22] that split satisfies $\texttt{split}(h, \langle a,b\rangle) = h(a,b)$. The equations for list_case follow easily by rewriting with the those for case and split.

$$\texttt{list_case}(c,h,\texttt{Nil}) = \texttt{case}(\lambda u\,.\,c, \texttt{split}(h), \texttt{Inl}(\emptyset))$$

$$= (\lambda u\,.\,c)(\emptyset)$$
$$= c\,.$$

$$\begin{aligned}\texttt{list_case}(c,h,\texttt{Cons}(x,y)) &= \texttt{case}(\lambda u\,.\,c,\texttt{split}(h),\texttt{Inr}(\langle x,y\rangle))\\ &= \texttt{split}(h,\langle x,y\rangle)\\ &= h(x,y).\end{aligned}$$

To summarize, we obtain the equations

$$\texttt{list_case}(c,h,\texttt{Nil}) = c \qquad \texttt{list_case}(c,h,\texttt{Cons}(x,y)) = h(x,y).$$

Proving the equations for `list_rec` is almost as easy. Unfolding the recursion equation for `Vrec` yields

$$\texttt{list_rec}(c,h,l) = \texttt{list_case}(c,\ \lambda x\,y\,.\,h(x,y,g`y),\ l) \tag{7}$$

where $g \equiv \lambda z \in V_{\texttt{rank}(l)}\,.\,\texttt{list_rec}(c,h,z)$. We instantly obtain the `Nil` case, and with slightly more effort, the recursive case:

$$\texttt{list_rec}(c,h,\texttt{Nil}) = c \qquad \texttt{list_rec}(c,h,\texttt{Cons}(x,y)) = h(x,y,\texttt{list_rec}(c,h,y))$$

In deriving the latter equation, the first step is to put $l \equiv \texttt{Cons}(x,y)$ in (7) and apply an equation for `list_case`:

$$\begin{aligned}\texttt{list_rec}(c,h,\texttt{Cons}(x,y)) &= \texttt{list_case}(c,\ \lambda x\,y\,.\,h(x,y,g`y),\texttt{Cons}(x,y))\\ &= h(x,y,g`y)\end{aligned}$$

All that remains is the $\beta$-reduction of $g`y$ to $\texttt{list_rec}(c,h,y)$, where $g`y$ is

$$(\lambda z \in V_{\texttt{rank}(\texttt{Cons}(x,y))}\,.\,\texttt{list_rec}(c,h,z))\ `\ y\,.$$

This step requires proving $y \in V_{\texttt{rank}(\texttt{Cons}(x,y))}$. Note that $\texttt{Cons}(x,y) = \langle 1,\langle x,y\rangle\rangle$; by properties of `rank` (§3.4), we must show

$$\texttt{rank}(y) < \texttt{rank}(\langle 1,\langle x,y\rangle\rangle).$$

This is obvious because $\texttt{rank}(b) < \texttt{rank}(\langle a,b\rangle)$ for all $a$ and $b$, and because the relation $<$ is transitive.

Recursion operators for other data structures are derived in the same manner.

4.3.2. *Defining Functions on Lists*

The Isabelle theory defines some common list operations, such as append and `map`, using `list_rec`:

$$\begin{aligned}\texttt{map}(h,l) &\equiv \texttt{list_rec}(\texttt{Nil},\ \lambda x\,y\,r\,.\,\texttt{Cons}(h(x),r),\ l)\\ xs@ys &\equiv \texttt{list_rec}(ys,\ \lambda x\,y\,r\,.\,\texttt{Cons}(x,r),\ xs)\end{aligned}$$

The usual recursion equations follow directly. Note the absence of typing conditions such as $l \in \texttt{list}(A)$:

$$\texttt{map}(h, \texttt{Nil}) = \texttt{Nil} \qquad \texttt{map}(h, \texttt{Cons}(a, l)) = \texttt{Cons}(h(a), \texttt{map}(h, l))$$
$$\texttt{Nil}@ys = ys \qquad \texttt{Cons}(a, l)@ys = \texttt{Cons}(a, l@ys)$$

The familiar theorems about these functions have elementary proofs by list induction and simplification. Theorems proved by induction have typing conditions; here is one example out of the many proved in Isabelle:

$$\frac{xs \in \texttt{list}(A)}{\texttt{map}(h, xs@ys) = \texttt{map}(h, xs)@\texttt{map}(h, ys)}$$

We can also prove some unusual type-checking rules:

$$\frac{l \in \texttt{list}(A)}{\texttt{map}(h, l) \in \texttt{list}(\{h(x) \,.\, x \in A\})}$$

Here, $\texttt{list}(\{h(x) \,.\, x \in A\})$ is the set of all lists whose elements have the form $h(x)$ for some $x \in A$. Using $\texttt{list}(\cdots)$ in recursive definitions raises interesting possibilities, as the next section will illustrate.

### 4.4. Using list(···) in Recursion Equations

Recursive data structure definitions typically involve $\times$ and $+$, but sometimes it is convenient to involve other set constructors. This section demonstrates using $\texttt{list}(\cdots)$ to define another data structure.

Consider the syntax of terms over the alphabet $A$. Each term is a function application $f(t_1, \ldots, t_n)$, where $f \in A$ and $t_1, \ldots, t_n$ are themselves terms. We shall formalize this syntax as $\texttt{term}(A)$, the set of all trees whose nodes are labelled with an element of $A$ and which have zero or more subtrees. It is natural to regard the subtrees as a list; we solve the recursion equation

$$\texttt{term}(A) = A \times \texttt{list}(\texttt{term}(A)). \tag{8}$$

Before using $\texttt{list}(\cdots)$ with the Knaster-Tarski Theorem, we must show that it is monotonic and bounded:

$$\frac{A \subseteq B}{\texttt{list}(A) \subseteq \texttt{list}(B)} \qquad \texttt{list}(\texttt{univ}(A)) \subseteq \texttt{univ}(A)$$

The proofs are simple using lemmas such as the monotonicity of lfp (§2.4). If we now define

$$\begin{aligned} \texttt{term}(A) &\equiv \texttt{lfp}(\texttt{univ}(A), \lambda X \,.\, A \times \texttt{list}(X)) \\ \texttt{Apply}(a, ts) &\equiv \langle a, ts \rangle \end{aligned}$$

then we quickly derive (8) and obtain the single introduction rule

$$\frac{a \in A \quad ts \in \texttt{list}(\texttt{term}(A))}{\texttt{Apply}(a, ts) \in \texttt{term}(A)}$$

The structural induction rule takes a curious form:

$$\frac{t \in \texttt{term}(A) \qquad \begin{matrix}[x \in A \quad zs \in \texttt{list}(\{z \in \texttt{term}(A)\,.\,\psi(z)\})]_{x,zs} \\ \vdots \\ \psi(\texttt{Apply}(x, zs))\end{matrix}}{\psi(t)}$$

Because of the presence of list in the recursion equation (8), we cannot express induction hypotheses in the familiar manner. Clearly, $zs \in \texttt{list}(\{z \in \texttt{term}(A)\,.\,\psi(z)\})$ if and only if every element $z$ of $zs$ satisfies $\psi(z)$ and belongs to $\texttt{term}(A)$. Proofs by this induction rule generally require a further induction over the term list $zs$.

### 4.4.1. *Recursion on Terms*

Let us define analogues of list_case and list_rec. The former is trivial: because every term is an ordered pair, we may use split.

A recursive function on terms will naturally apply itself to the list of subterms, using the list functional map. Define

$$\texttt{term_rec}(d, t) \;\equiv\; \texttt{Vrec}(t,\; \lambda t\, g\,.\,\texttt{split}(\lambda x\, zs\,.\, d(x, zs, \texttt{map}(\lambda z\,.\, g`z, zs)),\; t))$$

Note that map was defined above to be a binding operator; it applies to a meta-level function, not a ZF function (a set of pairs). Since $g$ denotes a ZF function, we must write $\texttt{map}(\lambda z\,.\, g`z, zs)$ instead of $\texttt{map}(g, zs)$. Although the form of map causes complications now, it leads to simpler equations later.

Put $t \equiv \texttt{Apply}(a, ts)$ in the definition of term_rec. Unfolding the recursion equation for Vrec and applying the equation for split yields

$$\begin{aligned}\texttt{term_rec}(d, \texttt{Apply}(a, ts)) &= \texttt{split}(\lambda x\, zs\,.\, d(x, zs, \texttt{map}(\lambda z\,.\, g`z, zs)),\; \langle a, ts\rangle) \\ &= d(a, ts, \texttt{map}(\lambda z\,.\, g`z, ts))\end{aligned}$$

where $g \equiv \lambda x \in V_{\texttt{rank}(\langle a,ts\rangle)}\,.\,\texttt{term_rec}(d, x)$. The map above applies $\texttt{term_rec}(d, x)$, restricted to $x$ such that $\texttt{rank}(x) < \texttt{rank}(\langle a, ts\rangle)$, to each member of $ts$. Clearly, each member of $ts$ has lesser rank than $ts$, and therefore lesser rank than $\langle a, ts\rangle$; the restriction on $x$ has no effect, and the result must equal $\texttt{map}(\lambda z.\texttt{term_rec}(d, z), ts)$. We may abbreviate this (by $\eta$-contraction) to $\texttt{map}(\texttt{term_rec}(d), ts)$.

To formalize this argument, the ZF theory proves the more general lemma

$$\frac{l \in \texttt{list}(A) \quad \texttt{Ord}(\alpha) \quad \texttt{rank}(l) \in \alpha}{\texttt{map}(\lambda z\,.\,(\lambda x \in V_\alpha\,.\,h(x))`z,\; l) = \texttt{map}(h, l)}$$

by structural induction on the list $l$. The lemma simplifies the term_rec equation to

$$\frac{ts \in \texttt{list}(A)}{\texttt{term_rec}(d, \texttt{Apply}(a, ts)) = d(a, ts, \texttt{map}(\texttt{term_rec}(d), ts))}$$

The curious premise $ts \in \texttt{list}(A)$ arises from the map lemma just proved; $A$ need not be a set of terms and does not appear in the conclusion. Possibly, this premise could be eliminated by reasoning about the result of map when applied to non-lists.

4.4.2. *Defining Functions on Terms*

To illustrate the use of term_rec, let us define the operation to reflect a term about its vertical axis, reversing the list of subtrees at each node. First we define rev, the traditional list reverse operation.[8]

$$\begin{aligned} \texttt{rev}(l) &\equiv \texttt{list_rec}(\texttt{Nil},\ \lambda x\, y\, r\, .\, r@\texttt{Cons}(x, r),\ l) \\ \texttt{reflect}(t) &\equiv \texttt{term_rec}(\lambda x\ zs\ rs\ .\ \texttt{Apply}(x, \texttt{rev}(rs)),\ t) \end{aligned}$$

Unfolding the recursion equation for term_rec instantly yields, for $ts \in \texttt{list}(A)$,

$$\texttt{reflect}(\texttt{Apply}(a, ts)) = \texttt{Apply}(a, \texttt{rev}(\texttt{map}(\texttt{reflect}, ts))). \tag{9}$$

Note the simple form of the map application above, since reflect is a meta-level function. Defining functions at the meta-level allows them to operate over the class of all sets. On the other hand, an object-level function is a set of pairs; its domain and range must be sets.

4.4.3. *An Induction Rule for Equations Between Terms*

The Isabelle ZF theory defines and proves theorems about several term operations. Many term operations involve a corresponding list operation, as reflect involves rev. Proofs by term induction involve reasoning about map.

Since many theorems are equations, let us derive an induction rule for proving equations easily. First, we derive two rules:

$$\frac{l \in \texttt{list}(\{x \in A\ .\ \psi(x)\})}{l \in \texttt{list}(A)} \qquad \frac{l \in \texttt{list}(\{x \in A\ .\ h_1(x) = h_2(x)\})}{\texttt{map}(h_1, l) = \texttt{map}(h_2, l)}$$

The first rule follows by monotonicity of list. To understand the second rule, suppose $l \in \texttt{list}(\{x \in A\ .\ h_1(x) = h_2(x)\})$. Then $h_1(x) = h_2(x)$ holds for every member $x$ of the list $l$, so $\texttt{map}(h_1, l) = \texttt{map}(h_2, l)$. This argument may be formalized using list induction.

Combining the two rules with term induction yields the derived induction rule:

$$\frac{t \in \texttt{term}(A) \qquad \begin{array}{c} [x \in A \quad zs \in \texttt{list}(\texttt{term}(A)) \quad \texttt{map}(h_1, zs) = \texttt{map}(h_2, zs)]_{x,zs} \\ \vdots \\ h_1(\texttt{Apply}(x, zs)) = h_2(\texttt{Apply}(x, zs)) \end{array}}{h_1(t) = h_2(t)}$$

The induction hypothesis, $\texttt{map}(h_1, zs) = \texttt{map}(h_2, zs)$, neatly expresses that $h_1(z) = h_2(z)$ holds for every member $z$ of the list $zs$.

4.4.4. *Example of Equational Induction*

To demonstrate the induction rule, let us prove $\texttt{reflect}(\texttt{reflect}(t)) = t$. The proof requires four lemmas about `rev` and `map`. Ignoring the premise $l \in \texttt{list}(A)$, the lemmas are

$$\texttt{rev}(\texttt{map}(h, l)) = \texttt{map}(h, \texttt{rev}(l)) \tag{10}$$

$$\texttt{map}(h_1, \texttt{map}(h_2, l)) = \texttt{map}(\lambda u \,.\, h_1(h_2(u)), l) \tag{11}$$

$$\texttt{map}(\lambda u \,.\, u, l) = l \tag{12}$$

$$\texttt{rev}(\texttt{rev}(l)) = l \tag{13}$$

To apply the derived induction rule, we may assume the induction hypothesis

$$\texttt{map}(\lambda u \,.\, \texttt{reflect}(\texttt{reflect}(u)), zs) = \texttt{map}(\lambda u \,.\, u, zs) \tag{14}$$

and must show

$$\texttt{reflect}(\texttt{reflect}(\texttt{Apply}(x, zs))) = \texttt{Apply}(x, zs).$$

Simplifying the left hand side, we have

$$\begin{aligned}
&\texttt{reflect}(\texttt{reflect}(\texttt{Apply}(x, zs))) \\
&= \texttt{reflect}(\texttt{Apply}(x, \texttt{rev}(\texttt{map}(\texttt{reflect}, zs)))) && \text{by (9)} \\
&= \texttt{reflect}(\texttt{Apply}(x, \texttt{map}(\texttt{reflect}, \texttt{rev}(zs)))) && \text{by (10)} \\
&= \texttt{Apply}(x, \texttt{rev}(\texttt{map}(\texttt{reflect}, \texttt{map}(\texttt{reflect}, \texttt{rev}(zs))))) && \text{by (9)} \\
&= \texttt{Apply}(x, \texttt{rev}(\texttt{map}(\lambda u \,.\, \texttt{reflect}(\texttt{reflect}(u)), \texttt{rev}(zs)))) && \text{by (11)} \\
&= \texttt{Apply}(x, \texttt{map}(\lambda u \,.\, \texttt{reflect}(\texttt{reflect}(u)), \texttt{rev}(\texttt{rev}(zs)))) && \text{by (10)} \\
&= \texttt{Apply}(x, \texttt{map}(\lambda u \,.\, \texttt{reflect}(\texttt{reflect}(u)), zs)) && \text{by (13)} \\
&= \texttt{Apply}(x, \texttt{map}(\lambda u \,.\, u, zs)) && \text{by (14)} \\
&= \texttt{Apply}(x, zs) && \text{by (12)}
\end{aligned}$$

The use of `map` may be elegant, but the proof is rather obscure. The next section describes an alternative formulation of the term data structure.

This section has illustrated how `list` can be added to our repertoire of set constructors permitted in recursive data structure definitions. It seems clear that other set constructors, including `term` itself, can be added similarly.

## 4.5. Mutual Recursion

Consider the sets $\texttt{tree}(A)$ and $\texttt{forest}(A)$ defined by the mutual recursion equations

$$\begin{aligned}
\texttt{tree}(A) &= A \times \texttt{forest}(A) \\
\texttt{forest}(A) &= \{\emptyset\} + \texttt{tree}(A) \times \texttt{forest}(A)
\end{aligned}$$

Observe that $\texttt{tree}(A)$ is essentially the same data structure as $\texttt{term}(A)$, since $\texttt{forest}(A)$ is essentially the same as $\texttt{list}(\texttt{term}(A))$. Mutual recursion avoids the complications of recursion over the operator `list`, but introduces its own complications.

### 4.5.1. *The General Approach*

Mutual recursion equations are typically solved by applying the Knaster-Tarski Theorem over the lattice $\wp(A)\times\wp(B)$, the Cartesian product of two powersets. But we have proved the Theorem only for a simple powerset lattice. Because the lattice $\wp(A+B)$ is order-isomorphic to $\wp(A)\times\wp(B)$, we shall instead apply the Theorem to a lattice of the form $\wp(A+B)$. We solve the equations by constructing a disjoint sum comprising all of the sets in the definition — here, a set called $\texttt{TF}(A)$, which will contain $\texttt{tree}(A)$ and $\texttt{forest}(A)$ as disjoint subsets. This approach appears to work well, and $\texttt{TF}(A)$ turns out to be useful in itself. A minor drawback: it does not solve the recursion equations up to equality, only up to isomorphism.

To support this approach to mutual recursion, define

$$\texttt{Part}(A,h) \equiv \{x \in A \,.\, \exists z \,.\, x = h(z)\}.$$

Here $\texttt{Part}(A,h)$ selects the subset of $A$ whose elements have the form $h(z)$. Typically $h$ is `Inl` or `Inr`, the injections for the disjoint sum. Note that $\texttt{Part}(A{+}B, \texttt{Inl})$ equals not $A$ but $\{\texttt{Inl}(x)\,.\,x\in A\}$. The disjoint sum of three or more sets involves nested injections. We may use `Part` with the composition of injections, such as $\lambda x\,.\,\texttt{Inr}(\texttt{Inl}(x))$, and obtain equations such as

$$\texttt{Part}(A+(B+C),\ \lambda x\,.\,\texttt{Inr}(\texttt{Inl}(x))) = \{\texttt{Inr}(\texttt{Inl}(x))\,.\,x\in B\}.$$

### 4.5.2. *The Formal Definitions*

Now $\texttt{TF}(A)$, $\texttt{tree}(A)$ and $\texttt{forest}(A)$ are defined by

$$\begin{aligned}
\texttt{TF}(A) &\equiv \texttt{lfp}(\texttt{univ}(A),\ \lambda X\,.\,A\times\texttt{Part}(X,\texttt{Inr}) + \\
&\qquad\qquad (\{\emptyset\} + \texttt{Part}(X,\texttt{Inl})\times\texttt{Part}(X,\texttt{Inr})))\\
\texttt{tree}(A) &\equiv \texttt{Part}(\texttt{TF}(A),\texttt{Inl})\\
\texttt{forest}(A) &\equiv \texttt{Part}(\texttt{TF}(A),\texttt{Inr})
\end{aligned}$$

The presence of `Part` does not complicate reasoning about `lfp`. In particular, $\texttt{Part}(A,h)$ is monotonic in $A$. We obtain

$$\begin{aligned}
\texttt{TF}(A) &= A\times\texttt{Part}(\texttt{TF}(A),\texttt{Inr}) + \\
&\quad(\{\emptyset\} + \texttt{Part}(\texttt{TF}(A),\texttt{Inl})\times\texttt{Part}(\texttt{TF}(A),\texttt{Inr}))\\
&= A\times\texttt{forest}(A) + (\{\emptyset\} + \texttt{tree}(A)\times\texttt{forest}(A))
\end{aligned}$$

This solves our recursion equations up to isomorphism:

$$\begin{aligned}
\texttt{tree}(A) &= \{\texttt{Inl}(x)\,.\,x\in A\times\texttt{forest}(A)\}\\
\texttt{forest}(A) &= \{\texttt{Inr}(x)\,.\,x\in\{\emptyset\}+\texttt{tree}(A)\times\texttt{forest}(A)\}
\end{aligned}$$

These equations determine the tree and forest constructors, Tcons, Fnil and Fcons. Due to the similarity to $\mathtt{list}(A)$, we can use the list constructors to abbreviate the definitions:

$$\begin{aligned}
\mathtt{Tcons}(a,f) &\equiv \mathtt{Inl}(\langle a,f\rangle) \\
\mathtt{Fnil} &\equiv \mathtt{Inr}(\mathtt{Nil}) \\
\mathtt{Fcons}(t,f) &\equiv \mathtt{Inr}(\mathtt{Cons}(t,f))
\end{aligned}$$

A little effort yields the introduction rules:

$$\frac{a\in A \quad f\in\mathtt{forest}(A)}{\mathtt{Tcons}(a,f)\in\mathtt{tree}(A)} \qquad \mathtt{Fnil}\in\mathtt{forest}(A) \qquad \frac{t\in\mathtt{tree}(A)\quad f\in\mathtt{forest}(A)}{\mathtt{Fcons}(t,f)\in\mathtt{forest}(A)}$$

The usual methods yield a structural induction rule for $\mathtt{TF}(A)$:

$$\frac{z\in\mathtt{TF}(A) \qquad \begin{array}{c}\left[\begin{array}{l}x\in A\\ f\in\mathtt{forest}(A)\\ \psi(f)\end{array}\right]_{x,f}\\ \vdots\\ \psi(\mathtt{Tcons}(x,f))\end{array} \qquad \psi(\mathtt{Fnil}) \qquad \begin{array}{c}\left[\begin{array}{l}t\in\mathtt{tree}(A)\\ f\in\mathtt{forest}(A)\\ \psi(t)\\ \psi(f)\end{array}\right]_{t,f}\\ \vdots\\ \psi(\mathtt{Fcons}(t,f))\end{array}}{\psi(z)} \tag{15}$$

(The assumptions are stacked vertically to save space.) Although this may not look like the best rule for mutual recursion, it is surprisingly simple and useful. It affords easy proofs of several theorems in the Isabelle theory. For the general case, there is a rule that allows different induction formulae, $\psi$ for trees and $\phi$ for forests:

$$\frac{\begin{array}{c}\left[\begin{array}{l}x\in A\\ f\in\mathtt{forest}(A)\\ \phi(f)\end{array}\right]_{x,f}\\ \vdots\\ \psi(\mathtt{Tcons}(x,f))\end{array} \qquad \phi(\mathtt{Fnil}) \qquad \begin{array}{c}\left[\begin{array}{l}t\in\mathtt{tree}(A)\\ f\in\mathtt{forest}(A)\\ \psi(t)\\ \phi(f)\end{array}\right]_{t,f}\\ \vdots\\ \phi(\mathtt{Fcons}(t,f))\end{array}}{(\forall_{t\in\mathtt{tree}(A)}\,.\,\psi(t))\wedge(\forall_{f\in\mathtt{forest}(A)}\,.\,\phi(f))} \tag{16}$$

This rule follows by applying the previous one to the formula

$$(z\in\mathtt{tree}(A)\rightarrow\psi(z))\wedge(z\in\mathtt{forest}(A)\rightarrow\phi(z)).$$

Its derivation relies on the disjointness of $\mathtt{tree}(A)$ and $\mathtt{forest}(A)$. Both rules are demonstrated below.

4.5.3. *Operating on Trees and Forests*

The case analysis operator is called `TF_case` and the recursion operator is called `TF_rec`:

$$\begin{aligned}\texttt{TF_case}(b,c,d,z) &\equiv \texttt{case}(\texttt{split}(b),\texttt{list_case}(c,d),z)\\ \texttt{TF_rec}(b,c,d,z) &\equiv \texttt{Vrec}(z,\ \lambda z\, r\,.\,\texttt{TF_case}(\lambda x\ f\,.\,b(x,f,r`f),\\ &\qquad\qquad c,\ \lambda t\ f\,.\,d(t,f,r`t,r`f),\ z))\end{aligned}$$

Note the use of the case analysis operators for disjoint sums (`case`), Cartesian products (`split`), and lists (`list_case`). Unfolding `Vrec`, we now derive the recursion rules, starting with the one for trees:

$$\begin{aligned}&\texttt{TF_rec}(b,c,d,\texttt{Tcons}(a,f))\\ &\quad= \texttt{TF_rec}(b,c,d,\texttt{Inl}(\langle a,f\rangle))\\ &\quad= \texttt{TF_case}(\lambda x\ f\,.\,b(x,f,r`f),\ c,\ \lambda t\ f\,.\,d(t,f,r`t,r`f),\ \texttt{Inl}(\langle a,f\rangle))\\ &\quad= \texttt{case}(\texttt{split}(\lambda x\ f\,.\,b(x,f,r`f)),\\ &\qquad\qquad \texttt{list_case}(c,\ \lambda t\ f\,.\,d(t,f,r`t,r`f)),\ \texttt{Inl}(\langle a,f\rangle))\\ &\quad= \texttt{split}(\lambda x\ f\,.\,b(x,f,r`f),\ \langle a,f\rangle)\\ &\quad= b(a,f,r`f)\end{aligned}$$

where $r \equiv \lambda x \in V_{\texttt{rank(Inl(}\langle a,f\rangle\texttt{))}}\,.\,\texttt{TF_rec}(b,c,d,x)$. The usual lemmas prove

$$\texttt{rank}(f) < \texttt{rank}(\texttt{Inl}(\langle a,f\rangle)),$$

allowing the $\beta$-reduction of $r`f$ to $b(a,f,\texttt{TF_rec}(b,c,d,f))$. The other recursion rules for `TF_rec` are derived similarly. To summarize, we have

$$\begin{aligned}\texttt{TF_rec}(b,c,d,\texttt{Tcons}(a,f)) &= b(a,f,\texttt{TF_rec}(b,c,d,f))\\ \texttt{TF_rec}(b,c,d,\texttt{Fnil}) &= c\\ \texttt{TF_rec}(b,c,d,\texttt{Fcons}(t,f)) &= d(t,f,\texttt{TF_rec}(b,c,d,t),\texttt{TF_rec}(b,c,d,f))\end{aligned}$$

4.5.4. *Defining Functions on Trees and Forests*

Some examples may be helpful. Here are three applications of `TF_rec`:

- `TF_map` applies an operation to every label of a tree.
- `TF_size` returns the number of labels in a tree.
- `TF_preorder` returns the labels as a list, in preorder.

Each operation is defined simultaneously for trees and forests:

$$\texttt{TF_map}(h,z) \equiv \texttt{TF_rec}(\lambda x\ f\ r\,.\,\texttt{Tcons}(h(x),r),$$

$$
\begin{aligned}
& \qquad\qquad \mathtt{Fnil}, \\
& \qquad\qquad \lambda t\ f\ r_1\ r_2 \,.\, \mathtt{Fcons}(r_1, r_2),\ z) \\
\mathtt{TF_size}(h, z) &\equiv \mathtt{TF_rec}(\lambda x\ f\ r \,.\, \mathtt{succ}(r), \\
& \qquad\qquad 0, \\
& \qquad\qquad \lambda t\ f\ r_1\ r_2 \,.\, r_1 \oplus r_2,\ z) \\
\mathtt{TF_preorder}(h, z) &\equiv \mathtt{TF_rec}(\lambda x\ f\ r \,.\, \mathtt{Cons}(x, r), \\
& \qquad\qquad \mathtt{Nil}, \\
& \qquad\qquad \lambda t\ f\ r_1\ r_2 \,.\, r_1 @ r_2,\ z)
\end{aligned}
$$

Here $\oplus$ is the addition operator for natural numbers. Recall that @ is the append operator for lists (§4.3).

Applying the `TF_rec` recursion equations to `TF_map` immediately yields

$$
\begin{aligned}
\mathtt{TF_map}(h, \mathtt{Tcons}(a, f)) &= \mathtt{Tcons}(h(a), \mathtt{TF_map}(h, f)) \\
\mathtt{TF_map}(h, \mathtt{Fnil}) &= \mathtt{Fnil} \\
\mathtt{TF_map}(h, \mathtt{Fcons}(t, f)) &= \mathtt{Fcons}(\mathtt{TF_map}(h, t), \mathtt{TF_map}(h, f))
\end{aligned}
$$

Many theorems can be proved by the simple induction rule (15) for $\mathtt{TF}(A)$, taking advantage of ZF's lack of a formal type system. Separate proofs for $\mathtt{tree}(A)$ and $\mathtt{forest}(A)$ would require the cumbersome rule for mutual induction.

4.5.5. *Example of Simple Induction*

Let us prove $\mathtt{TF_map}(\lambda u \,.\, u, z) = z$ for all $z \in \mathtt{TF}(A)$. By the simple induction rule (15), it suffices to prove three subgoals:

- $\mathtt{TF_map}(\lambda u \,.\, u, \mathtt{Tcons}(x, f)) = \mathtt{Tcons}(x, f)$ assuming the induction hypothesis $\mathtt{TF_map}(\lambda u \,.\, u, f) = f$
- $\mathtt{TF_map}(\lambda u \,.\, u, \mathtt{Fnil}) = \mathtt{Fnil}$
- $\mathtt{TF_map}(\lambda u \,.\, u, \mathtt{Fcons}(t, f)) = \mathtt{Fcons}(t, f)$ assuming the induction hypotheses $\mathtt{TF_map}(\lambda u \,.\, u, t) = t$ and $\mathtt{TF_map}(\lambda u \,.\, u, f) = f$

These are all trivial, by the recursion equations. For example, the first subgoal is proved in two steps:

$$\mathtt{TF_map}(\lambda u \,.\, u, \mathtt{Tcons}(x, f)) = \mathtt{Tcons}(x, \mathtt{TF_map}(\lambda u \,.\, u, f)) = \mathtt{Tcons}(x, f)$$

The simple induction rule proves various laws relating `TF_map`, `TF_size` and `TF_preorder` with equal ease.

4.5.6. *Example of Mutual Induction*
The mutual induction rule (16) proves separate properties for $\texttt{tree}(A)$ and $\texttt{forest}(A)$. The simple rule (15) can show that $\texttt{TF_map}$ takes elements of $\texttt{TF}(A)$ to $\texttt{TF}(B)$, for some $B$; let us sharpen this result to show that $\texttt{TF_map}$ takes trees to trees and forests to forests. Assume $h(x) \in B$ for all $x \in A$ and apply mutual induction to the formula

$$(\forall_{t\in\texttt{tree}(A)} . \texttt{TF_map}(h,t) \in \texttt{tree}(B)) \wedge (\forall_{f\in\texttt{forest}(A)} . \texttt{TF_map}(h,f) \in \texttt{forest}(B))$$

The first subgoal of the induction is to show

$$\texttt{TF_map}(h, \texttt{Tcons}(x, f)) \in \texttt{tree}(B)$$

assuming $x \in A$, $f \in \texttt{forest}(A)$ and $\texttt{TF_map}(h,f) \in \texttt{forest}(B)$. The recursion equation for $\texttt{TF_map}$ reduces it to

$$\texttt{Tcons}(h(x), \texttt{TF_map}(h,f)) \in \texttt{tree}(B);$$

the type-checking rules for $\texttt{Tcons}$ and $h$ reduce it to the assumptions $x \in A$ and $\texttt{TF_map}(h,f) \in \texttt{forest}(B)$.

The second subgoal of the induction is

$$\texttt{TF_map}(h, \texttt{Fnil}) \in \texttt{forest}(B),$$

which reduces to the trivial $\texttt{Fnil} \in \texttt{forest}(B)$. The third subgoal,

$$\texttt{TF_map}(h, \texttt{Fcons}(t, f)) \in \texttt{forest}(B),$$

is treated like the first.

We have considered two approaches to defining variable-branching trees. The previous section defines $\texttt{term}(A)$ by recursion over the operator $\texttt{list}$, so that $\texttt{list}(\texttt{term}(A))$ denotes the set of forests over $A$. I prefer this to the present approach of mutual recursion. But this one example does not demonstrate that mutual recursion should always be avoided. An example to study is a programming language that allows embedded commands in expressions; its expressions and commands would be mutually recursive.

## 5. Soundness and Completeness of Propositional Logic

We have discussed the ZF formalization of least fixedpoints, recursive functions and recursive data structures. Formalizing propositional logic — its syntax, semantics and proof theory — exercises each of these principles. The proofs of soundness and completeness amount to an equivalence proof between denotational and operational semantic definitions. Similar examples abound in theoretical Computer Science.

### 5.1. Defining the Set of Propositions

The *propositions* come in three forms:

1. $\texttt{Fls}$ is the absurd proposition.
2. $\#v$ is a propositional variable, for $v \in \texttt{nat}$.
3. $p \supset q$ is an implication if $p$ and $q$ are propositions.

The set $\texttt{prop}$ consists of all propositions. It is the least solution to the recursion equation

$$\texttt{prop} = \{\emptyset\} + \texttt{nat} + \texttt{prop} \times \texttt{prop}.$$

The definition is similar to the others described above. We obtain the introduction rules

$$\texttt{Fls} \in \texttt{prop} \qquad \frac{v \in \texttt{nat}}{\#v \in \texttt{prop}} \qquad \frac{p \in \texttt{prop} \quad q \in \texttt{prop}}{p \supset q \in \texttt{prop}}$$

with the usual induction rule for proving a property for every element of $\texttt{prop}$. Recursive functions on $\texttt{prop}$ are defined in the standard way.

Next, we define the denotational semantics of a proposition by translation to first-order logic. A *truth valuation* $t$ is a subset of $\texttt{nat}$ representing a set of atoms regarded as true (all others to be regarded as false). If $p \in \texttt{prop}$ and $t \subseteq \texttt{nat}$ then $\texttt{is_true}(p, t)$ states that $p$ evaluates to true under $t$. Writing $\bot$ for the absurd formula in first-order logic, the recursion equations are

$$\begin{aligned} \texttt{is_true}(\texttt{Fls}, t) &\leftrightarrow \bot \\ \texttt{is_true}(\#v, t) &\leftrightarrow v \in t \\ \texttt{is_true}(p \supset q, t) &\leftrightarrow (\texttt{is_true}(p, t) \to \texttt{is_true}(q, t)) \end{aligned}$$

Our recursion principles cannot express $\texttt{is_true}(p, t)$ directly since it is a formula. Instead, $\texttt{is_true}(p, t)$ is defined in terms of a recursive function that yields the truth value of $p$ as an element of $\{0, 1\}$. The details are omitted.

### 5.2. Defining an Inference System in ZF

Let $H$ be a set of propositions and $p$ a proposition. Write $H \models p$ to mean that the truth of all elements of $H$ implies the truth of $p$, for every truth valuation $t$. *Logical consequence* is formalized in ZF by

$$H \models p \equiv \forall t \,.\, (\forall_{q \in H} \,.\, \texttt{is_true}(q, t)) \to \texttt{is_true}(p, t)$$

The objective is to prove that $H \models p$ holds if and only if $p$ is provable from $H$ using the axioms $(K)$, $(S)$, $(DN)$ with the Modus Ponens rule $(MP)$. Note that $\supset$ associates to the right:

$$p \supset q \supset p \tag{$K$}$$

$$(p \supset q \supset r) \supset (p \supset q) \supset (p \supset r) \qquad (S)$$

$$((p \supset \texttt{Fls}) \supset \texttt{Fls}) \supset p \qquad (DN)$$

$$\frac{p \supset q \quad p}{q} \qquad (MP)$$

Such inference systems are becoming popular for defining the operational semantics of programming languages. They can be extremely large — consider the Definition of Standard ML [17]. The Knaster-Tarski Theorem can express the least set of propositions closed under the axioms and rules, but we must adopt a formalization that scales up to large inference systems.

Defining a separate Isabelle constant for each axiom and rule affords some control over formula expansion during proof. An axiom is expressed as a union over its schematic variables:

$$\begin{aligned}
\texttt{axK} &\equiv \bigcup_{p\in\texttt{prop}} \bigcup_{q\in\texttt{prop}} \{p \supset q \supset p\} \\
\texttt{axS} &\equiv \bigcup_{p\in\texttt{prop}} \bigcup_{q\in\texttt{prop}} \bigcup_{r\in\texttt{prop}} \{(p \supset q \supset r) \supset (p \supset q) \supset (p \supset r)\} \\
\texttt{axDN} &\equiv \bigcup_{p\in\texttt{prop}} \{((p \supset \texttt{Fls}) \supset \texttt{Fls}) \supset p\}
\end{aligned}$$

A rule takes a set $X$ of theorems and generates the set of all immediate consequences of $X$:

$$\texttt{ruleMP}(X) \equiv \bigcup_{p\in\texttt{prop}} \{q \in \texttt{prop} \,.\, \{p \supset q, p\} \subseteq X\}$$

The axioms and rules could have been defined in many equivalent ways. Unions and singletons give a uniform format for the axioms. But `ruleMP` makes an ad-hoc use of the Axiom of Separation, since its conclusion is just a schematic variable; this need not be the case for other rules. The use of the subset relation in $\{p \supset q, p\} \subseteq X$ simplifies the proof that $\texttt{ruleMP}(X)$ is monotonic in $X$.

We now define the set $\texttt{thms}(H)$ of theorems provable from $H$, and the consequence relation $H \vdash p$. The first part of the union, $H \cap \texttt{prop}$, considers only the *propositions* in $H$ as theorems; putting just $H$ here would make most of our results conditional on $H \subseteq \texttt{prop}$.

$$\begin{aligned}
\texttt{thms}(H) &\equiv \texttt{lfp}(\texttt{prop}, \ \lambda X \,.\, (H \cap \texttt{prop}) \cup \texttt{axK} \cup \texttt{axS} \cup \texttt{axDN} \cup \texttt{ruleMP}(X)) \\
H \vdash p &\equiv p \in \texttt{thms}(H)
\end{aligned}$$

We immediately obtain introduction rules corresponding to the axioms; the premises perform type-checking:

$$\frac{p \in H \quad p \in \texttt{prop}}{H \vdash p}\ (H) \qquad \frac{p \in \texttt{prop} \quad q \in \texttt{prop}}{H \vdash p \supset q \supset p}\ (K)$$

$$\frac{p \in \texttt{prop} \quad q \in \texttt{prop} \quad r \in \texttt{prop}}{H \vdash (p \supset q \supset r) \supset (p \supset q) \supset (p \supset r)}\ (S) \qquad \frac{p \in \texttt{prop}}{H \vdash ((p \supset \texttt{Fls}) \supset \texttt{Fls}) \supset p}\ (DN)$$

Proving that every theorem is a proposition helps to derive a rule for Modus Ponens that is free of type-checking:

$$\frac{H \vdash p}{p \in \texttt{prop}} \qquad \frac{H \vdash p \supset q \quad H \vdash p}{H \vdash q}\ (MP)$$

We may use these rules, cumbersome though they are, as an Isabelle object-logic. They can be supplied to tools such as the classical reasoner in order to prove Isabelle goals involving assertions of the form $H \vdash p$. This rule is derived using $(MP)$, $(S)$ and $(K)$:

$$\frac{p \in \texttt{prop}}{H \vdash p \supset p}\ (I)$$

By the monotonicity result from §2.4, $\texttt{thms}(H)$ is monotonic in $H$, which justifies a rule for weakening on the left. Axiom $(K)$ justifies weakening on the right:

$$\frac{G \subseteq H \quad G \vdash p}{H \vdash p} \qquad \frac{H \vdash q \quad p \in \texttt{prop}}{H \vdash p \supset q}$$

## 5.3. Rule Induction

Because it is defined using a least fixedpoint in ZF, our propositional logic admits induction over its proofs. This principle, sometimes called *rule induction*, does not require an explicit data structure for proofs; just apply the usual induction rule for `lfp`. Below we shall discuss this rule with two examples of its use, the Deduction Theorem and the Soundness Theorem (proving the latter in an Isabelle session).

The rule is too large to display in the usual notation. Its conclusion is $\psi(p)$ and it has six premises:

1. $H \vdash p$, which is the major premise
2. $\psi(x)$ with assumptions $[x \in \texttt{prop} \quad x \in H]_x$
3. $\psi(x \supset y \supset x)$ with assumptions $[x \in \texttt{prop} \quad y \in \texttt{prop}]_{x,y}$
4. $\psi((x \supset y \supset z) \supset (x \supset y) \supset x \supset z)$
   with assumptions $[x \in \texttt{prop} \quad y \in \texttt{prop} \quad z \in \texttt{prop}]_{x,y,z}$
5. $\psi(((x \supset \texttt{Fls}) \supset \texttt{Fls}) \supset x)$ with assumption $[x \in \texttt{prop}]_x$
6. $\psi(y)$ with assumptions $[H \vdash x \supset y \quad H \vdash x \quad \psi(x \supset y) \quad \psi(x)]_{x,y}$

The rationale for this form of induction is simple: if $\psi$ holds for all the axioms and is preserved by all the rules, then it must hold for all the theorems. The premise $\psi(x \supset y \supset x)$ ensures that $\psi$ holds for all instances of axiom $(K)$, and similar premises handle the other axioms. The last premise ensures that rule $(MP)$ preserves $\psi$; thus it takes $\psi(x \supset y)$ and $\psi(x)$ as induction hypotheses.[9]

The Deduction Theorem states that $\{p\} \cup H \vdash q$ implies $H \vdash p \supset q$. In Isabelle's set theory, it is formalized as follows (since $\mathtt{cons}(p, H) = \{p\} \cup H$):

$$\frac{\mathtt{cons}(p, H) \vdash q \quad p \in \mathtt{prop}}{H \vdash p \supset q}$$

The proof is by rule induction on $\mathtt{cons}(p, H) \vdash q$. Of the five remaining subgoals, the first is to show $H \vdash p \supset x$ assuming $x \in \mathtt{prop}$ and $x \in \mathtt{cons}(p, H)$. From $x \in \mathtt{cons}(p, H)$ there are two subcases:

- If $x = p$ then $H \vdash x \supset x$ follows using $(I)$.
- If $x \in H$ then $H \vdash p \supset x$ follows using $(H)$ and weakening.

The next three subgoals correspond to one of the axioms $(K)$, $(S)$ or $(DN)$, and hold by that axiom plus weakening. For the last subgoal, $H \vdash p \supset y$ follows from $H \vdash p \supset x \supset y$ and $H \vdash p \supset x$ using $(S)$ and $(MP)$.

Isabelle executes this proof of the Deduction Theorem in under six seconds. The classical reasoner, given the relevant lemmas, proves each subgoal automatically.

## 5.4. Proving the Soundness Theorem in Isabelle

Another application of rule induction is the Soundness Theorem:

$$\frac{H \vdash p}{H \models p}$$

The proof is straightforward. The most difficult case is showing that $H \models x \supset y$ and $H \models x$ imply $H \models y$. The Isabelle proof consists of three tactics. The `goalw` command states the goal and expands the definition of logical consequence, `logcon_def`.

```
goalw PropThms.thy [logcon_def] "!!H. H |- p ==> H |= p";
 Level 0
 !!H. H |- p ==> H |= p
  1. !!H. H |- p ==> ALL t. (ALL q:H. is_true(q, t)) --> is_true(p, t)
```

Applying rule induction to the premise $H \vdash p$ returns five subgoals:

```
by (eresolve_tac [PropThms.induct] 1);
```

```
Level 1
!!H. H |- p ==> H |= p
 1. !!H p.
       [| p : H; p : prop |] ==>
       ALL t. (ALL q:H. is_true(q, t)) --> is_true(p, t)
 2. !!H p q.
       [| p : prop; q : prop |] ==>
       ALL t. (ALL q:H. is_true(q, t)) --> is_true(p => q => p, t)
 3. !!H p q r.
       [| p : prop; q : prop; r : prop |] ==>
       ALL t.
          (ALL q:H. is_true(q, t)) -->
          is_true((p => q => r) => (p => q) => p => r, t)
 4. !!H p.
       p : prop ==>
       ALL t.
          (ALL q:H. is_true(q, t)) -->
          is_true(((p => Fls) => Fls) => p, t)
 5. !!H p q.
       [| H |- p => q;
          ALL t. (ALL q:H. is_true(q, t)) --> is_true(p => q, t);
          H |- p; ALL t. (ALL q:H. is_true(q, t)) --> is_true(p, t);
          p : prop; q : prop |] ==>
       ALL t. (ALL q:H. is_true(q, t)) --> is_true(q, t)
```

The equations for `is_true`, shown in §5.1 above, are called `is_true_Fls`, `is_true_Var` and `is_true_Imp` in Isabelle. Each is an 'if and only if' assertion. The next command converts `is_true_Imp` into the rule

$$\frac{\texttt{is_true}(p \supset q, t) \quad \texttt{is_true}(p, t)}{\texttt{is_true}(q, t)}$$

and gives it to `fast_tac`. The rule breaks down an induction hypothesis to solve subgoal 5.

```
by (fast_tac (ZF_cs addSDs [is_true_Imp RS iffD1 RS mp]) 5);
 Level 2
 !!H. H |- p ==> H |= p
   As above but without subgoal 5...
```

Rewriting by the recursion equations for `is_true`, Isabelle's simplifier solves the other four subgoals. For example, the conclusion of subgoal 2 rewrites to

$$\texttt{is_true}(x, t) \to \texttt{is_true}(y, t) \to \texttt{is_true}(x, t),$$

which is obviously true.

```
by (ALLGOALS
    (simp_tac
     (ZF_ss addsimps [is_true_Fls, is_true_Var, is_true_Imp])));
 Level 3
 !!H. H |- p ==> H |= p
 No subgoals!
```

This proof executes in about six seconds.

### 5.5. Completeness

Completeness means every valid proposition is provable: if $H \models p$ then $H \vdash p$. We consider first the special case where $H = \emptyset$ and later generalize $H$ to be any finite set.

A key lemma is the Law of the Excluded Middle, '$q$ or not $q$.' Since our propositions lack a disjunction symbol, the Law is expressed as a rule that reduces $p$ to two subgoals — one assuming $q$ and one assuming $\neg q$:

$$\frac{\texttt{cons}(q, H) \vdash p \quad \texttt{cons}(q \supset \texttt{Fls}, H) \vdash p \quad q \in \texttt{prop}}{H \vdash p}$$

#### 5.5.1. *The Informal Proof*

Let $t$ be a truth valuation and define $\texttt{hyps}(p, t)$ by recursion on $p$:

$$\begin{aligned} \texttt{hyps}(\texttt{Fls}, t) &= \emptyset \\ \texttt{hyps}(\#v, t) &= \begin{cases} \{\#v\} & \text{if } v \in t \\ \{\#v \supset \texttt{Fls}\} & \text{if } v \notin t \end{cases} \\ \texttt{hyps}(p \supset q, t) &= \texttt{hyps}(p, t) \cup \texttt{hyps}(q, t) \end{aligned}$$

Informally, $\texttt{hyps}(p, t)$ returns a set containing each atom in $p$, or the negation of that atom, depending on its value in $t$. The set $\texttt{hyps}(p, t)$ is necessarily finite.

For this section, call $H$ a *basis* of $p$ if $H \vdash p$. Assume that $p$ is valid, $\emptyset \models p$. After proving a lemma by induction, we find that $\texttt{hyps}(p, t)$ is a basis of $p$ for every truth valuation $t$:

$$\frac{p \in \texttt{prop} \quad \emptyset \models p}{\texttt{hyps}(p, t) \vdash p}$$

The next step towards establishing $\emptyset \vdash p$ is to reduce the size of the basis. If $\texttt{hyps}(p, t) = \texttt{cons}(\#v, H)$, then the basis contains $\#v$; removing $v$ from $t$ creates an almost identical basis that contains $\neg\#v$:

$$\texttt{hyps}(p, t - \{v\}) = \texttt{cons}(\#v \supset \texttt{Fls}, H) - \{\#v\}.$$

Applying the Law of the Excluded Middle with $\#v$ for $q$ yields $H \vdash p$, which is a basis of $p$ not mentioning $\#v$ at all. Repeating this operation yields smaller and smaller bases of $p$. Since $\texttt{hyps}(p,t)$ is finite, the empty set is also a basis. Thus we obtain $\emptyset \vdash p$, as desired.

5.5.2. *An Inductive Definition of Finite Sets*

The formalization of this argument is complex and will be omitted here. But one detail is relevant to recursive definitions: what is a finite set? Finite sets could be defined by reference to the natural numbers, but they are more easily defined as a least fixedpoint. The empty set is finite; if $y$ is finite then $\texttt{cons}(x,y)$ is also:

$$\texttt{Fin}(A) \equiv \texttt{lfp}(\wp(A),\ \lambda Z\,.\,\{\emptyset\} \cup (\bigcup_{y\in Z}\bigcup_{x\in A}\{\texttt{cons}(x,y)\}))$$

Monotonicity is shown by the usual lemmas; the Knaster-Tarski Theorem immediately yields the introduction rules:

$$\{\emptyset\} \in \texttt{Fin}(A) \qquad \frac{a \in A \quad b \in \texttt{Fin}(A)}{\texttt{cons}(a,b) \in \texttt{Fin}(A)}$$

We have defined a finite powerset operator; $\texttt{Fin}(A)$ consists of all the finite subsets of $A$. The induction rule for $\texttt{Fin}(A)$ resembles the rule for lists:

$$\frac{b \in \texttt{Fin}(A) \quad \psi(\emptyset) \quad \begin{matrix}[x \in A \quad y \in \texttt{Fin}(A) \quad x \notin y \quad \psi(y)]_{x,y} \\ \vdots \\ \psi(\texttt{cons}(x,y))\end{matrix}}{\psi(b)}$$

This rule strengthens the usual assumption rule for $\texttt{lfp}$ by discharging the assumption $x \notin y$. Its proof notes that $x \in y$ implies $\texttt{cons}(x,y) = y$, rendering the induction step trivial in this case.

Reasoning about finiteness is notoriously tricky, but finite set induction proves many results about $\texttt{Fin}(A)$ easily. The union of two finite sets is finite; the union of a finite set of finite sets is finite; a subset of a finite set is finite:

$$\frac{b \in \texttt{Fin}(A) \quad c \in \texttt{Fin}(A)}{b \cup c \in \texttt{Fin}(A)} \qquad \frac{C \in \texttt{Fin}(\texttt{Fin}(A))}{\bigcup C \in \texttt{Fin}(A)} \qquad \frac{c \subseteq b \quad b \in \texttt{Fin}(A)}{c \in \texttt{Fin}(A)}$$

5.5.3. *The Variable-Elimination Argument*

Returning to the completeness theorem, we can now prove that $\texttt{hyps}(p,t)$ is finite by structural induction on $p$:

$$\frac{p \in \texttt{prop}}{\texttt{hyps}(p,t) \in \texttt{Fin}(\bigcup_{v\in\texttt{nat}} .\{\#v, \#v \supset \texttt{Fls}\})}$$

For the variable-elimination argument, we assume $p \in \texttt{prop}$ and $\emptyset \models p$, and prove

$$\forall t\,.\,\texttt{hyps}(p,t) - \texttt{hyps}(p,t_0) \vdash p$$

by induction on the finite set $\mathtt{hyps}(p, t_0)$. (Here $t_0$ is simply a free variable.) Finally, instantiating $t$ to $t_0$ and using $A - A = \emptyset$, we obtain $\emptyset \vdash p$.

This establishes an instance of the Completeness Theorem:

$$\frac{\emptyset \models p \quad p \in \mathtt{prop}}{\emptyset \vdash p}$$

To show $H \models p$ implies $H \vdash p$ where $H$ may be any finite set requires a further application of finite set induction. I have not considered the case where $H$ is infinite, since it seems irrelevant to computational reasoning.

## 6. Related Work and Conclusions

This theory is intended to support machine proofs about recursive definitions. Every set theorist knows that ZF can handle recursion in principle, but machine proofs require assertions to be formalized correctly and conveniently. The derivations of the recursion operators `wfrec`, `transrec` and `Vrec` are particularly sensitive to formal details. Let us recall the chief problems, and their solutions:

- *Inductively defined sets* are expressed as least fixedpoints, applying the Knaster-Tarski Theorem over a suitable set.
- *Recursive functions* are defined by well-founded recursion and its derivatives, such as transfinite recursion.
- *Recursive data structures* are expressed by applying the Knaster-Tarski Theorem to a set with strong closure properties.

I have not attempted to characterize the class of recursive definitions admitted by these methods, but they are extremely general.

The overall approach is not restricted to ZF set theory. I have applied it, with a few changes, to Isabelle's implementation of higher-order logic. It may be applicable to weaker systems such as intuitionistic second-order logic and intuitionistic ZF set theory. Thus, we have a generic treatment of recursion for generic theorem proving.

In related work, Noël [18] has proved many theorems about recursion using Isabelle's set theory, including well-founded recursion and a definition of lists. But Noël does not develop a general theory of recursion. Ontic [10] provides strong support for recursively defined functions and sets. Ontic's theory of recursion differs from mine; it treats recursive functions as least fixedpoints, with no use of well-founded relations.

The Knaster-Tarski Theorem can be dropped. If $h$ is continuous then $\bigcup_{n\in\omega} h^n(\emptyset)$ is its least fixedpoint. Induction upon $n$ yields *computation induction*, which permits reasoning about the least fixedpoint. Ontic and Noël both use the construction, which generalizes to larger ordinals, but I have used it only to define `univ` and `eclose`.

The Knaster-Tarski Theorem has further applications in its dual form, which yields greatest fixedpoints. These crop up frequently in Computer Science, mainly in connection with bisimulation proofs [16].

Recently I have written an ML package to automate recursive definitions in Isabelle ZF [24]. My package is inspired by T. Melham's inductive definition packages for the Cambridge HOL system [5, 15]. It is unusually flexible because of its explicit use of the Knaster-Tarski Theorem. Monotone operators may occur in the introduction rules, such as the occurrence of `list` in the definition of $\mathtt{term}(A)$ above.

Given the desired form of the introduction rules, my package makes fixedpoint definitions. Then it proves the introduction and induction rules. It can define the constructors for a recursive data structure and prove their freeness. The package has been applied to most of the inductive definitions presented in this paper. It supports inductively defined relations and mutual recursion.

The Isabelle ZF theory described in this paper is available by ftp. For more information, please send electronic mail to the author, `lcp@cl.cam.ac.uk`.

## Acknowledgements

Martin Coen, Sara Kalvala and Philippe Noël commented on this paper. Tobias Nipkow (using Isabelle's higher-order logic) contributed the propositional logic example of §5. Thomas Melham suggested defining the finite powerset operator. Thanks are also due to Deepak Kapur (the editor) and to the four referees.

The research was funded by the EPSRC (grants GR/G53279, GR/H40570) and by the ESPRIT Basic Research Actions 3245 'Logical Frameworks' and 6453 'Types.'

## Notes

[1] This means the two sets are in one-to-one correspondence and have equivalent orderings.

[2] The `bnd_mono` premises could be weakened, but to little purpose, because they hold in typical uses of `lfp`.

[3] All Isabelle timings are on a Sun SPARCstation ELC.

[4] The approach could be generalized to non-well-founded set theory [2] by verifying that the set $\mathtt{univ}(A)$, defined in §4.2, is well-founded.

[5] There is no loss of generality: you can always apply transitive closure again.

[6] The traditional Axiom of Infinity has an existentially quantified variable in place of `Inf`. Introducing the constant is conservative, and allows `nat` to be defined explicitly.

[7] Earlier versions of Isabelle ZF defined $\mathtt{list}(A)$ to satisfy the recursion $\mathtt{list}(A) = \{\emptyset\} \cup (A \times \mathtt{list}(A))$. Then $\emptyset$ stood for the empty list and $\langle a, l\rangle$ for the list with head $a$ and tail $l$; note that $\emptyset$ does not equal any pair. The present approach follows a uniform treatment of data structures.

[8] This version takes quadratic time but it is easier to reason about than a linear time reverse.

[9] The other hypotheses, $H \vdash x \supset y$ and $H \vdash x$, are typical of *strong* rule induction [5]; they come for free from the induction rule for `lfp`.